\documentclass[11pt]{article}

\usepackage[margin=1in]{geometry}
\usepackage{setspace}
\usepackage{graphicx} % Required for inserting images
\usepackage{amsmath,amssymb} % Mathematical symbols
\usepackage{amsfonts}
\usepackage{booktabs}
\usepackage{xcolor}
\usepackage{appendix}
\usepackage{subcaption}

\usepackage{xurl}
\usepackage{hyperref}
\hypersetup{
    colorlinks=true,
    urlcolor=blue,
    linkcolor=blue,
    citecolor=blue
}
\usepackage[
    sorting=none,
    style=phys,
    backend=biber
]{biblatex}
\graphicspath{{figures/}}

\title{Activity-dependent epidemic spreading on multiscale brain networks predicts Alzheimer's disease progression}
\author{
Christoffer G. Alexandersen$^{1,2}$,
Suman S. Kulkarni$^{3}$,
Jessica T. Davis$^{4}$,\\
Sebastian N. Roemer-Cassiano$^{5}$,
Nicolai Franzmeier$^{5}$,
and Dani S. Bassett$^{2,6,7}$
}

\date{
$^{1}$Department of Bioengineering, University of Pennsylvania, Philadelphia, PA, USA\\
$^{2}$Wu Tsai Institute, Yale University, New Haven, CT, USA\\
$^{3}$Department of Physics, University of Pennsylvania, Philadelphia, PA, USA\\
$^{4}$Network Science Institute, Northeastern University, Boston, MA, USA\\
$^{5}$Institute for Stroke and Dementia Research, University Hospital,
Ludwig-Maximilians-Universität, Munich, Germany \\
$^{6}$Department of Psychology, Yale University, New Haven, CT USA\\
$^{7}$Department of Biomedical Engineering, Yale University, New Haven, CT, USA\\
}

\begin{document}

\maketitle

\begin{abstract}
    \noindent Neurodegenerative diseases can be viewed as spreading processes on brain networks, in which pathological proteins propagate between anatomically connected brain regions. Mathematical models have been used to study this process, but they generally ignore the influence of neuronal activity, even though experimental studies show that neuronal firing promotes protein transmission. Here, we couple a general node-activity process to susceptible--infected--susceptible dynamics. In this framework, an epidemic threshold determines whether small pathological seeds can grow, while a dominant network mode determines where growth begins. We derive approximations showing how neuronal activity shifts this threshold and redirects spreading by mixing structural network modes. For networks with multiscale structure, we decompose these changes into contributions from regional mean activity and within-region activity variation, allowing us to account for activity heterogeneity that is not resolved by brain imaging. Stochastic simulations validate the theoretical results across synthetic networks. We next use longitudinal human positron emission tomography to test whether neuronal activity predicts where and how broadly pathology spreads. Regional glucose metabolism serves as a proxy for neuronal activity, while tau accumulation measures disease progression. Adding neuronal activity to the network model captures spatial patterns of disease progression that are not explained by structural connectivity and established disease markers alone. Across individuals, predicted epidemic thresholds are also associated with how broadly pathology spreads through the brain. Together, these results connect epidemic theory to neurodegeneration, implicate neuronal activity as a driver of Alzheimer's disease progression, and motivate activity-modulating therapies to slow or prevent pathological spread.
\end{abstract}

\newpage
\section{Introduction}

Neurodegenerative diseases can be viewed as spreading processes on brain networks, in which pathological proteins propagate between anatomically connected regions and contribute to neuronal dysfunction, cell death, and cognitive decline~\cite{liu_trans-synaptic_2012,ahmed_novel_2014,raj_network_2012,zhou_predicting_2012,vogel_spread_2020,franzmeier_functional_2020}. Alzheimer's disease provides a prominent example through tau, a neuronal protein that accumulates abnormally and spreads through the brain as disease progresses. Tau pathology can remain largely confined to the medial temporal lobe during ageing, whereas Alzheimer's disease is associated with its widespread progression into the neocortex~\cite{braak_neuropathological_1991,scholl_pet_2016,crary_primary_2014}. Understanding disease progression therefore requires explaining both the spatial pattern of pathological spread and why localized pathology sometimes becomes widespread.

Because pathological proteins preferentially propagate between connected brain regions, mathematical models have represented neurodegenerative disease progression as spreading on the structural connectome. These models reproduce important spatial and temporal patterns of pathology~\cite{raj_network_2012,yang_longitudinal_2019,yang_longitudinal_2021,vogel_spread_2020,schafer_network_2020,schafer_predicting_2021,chaggar_personalised_2025,fornari_prion-like_2019}. However, they generally determine transmission from anatomy alone. Structural connectivity specifies the routes along which pathology can spread, but biological processes may strengthen or weaken transmission along those routes. Omitting these processes may therefore limit how well models explain both disease progression and disease onset.

Neuronal activity is one biological process that may modify pathological transmission through the structural network. Experimental studies show that neuronal firing increases protein release and transneuronal tau propagation~\cite{pooler_physiological_2013,yamada_neuronal_2014,wu_neuronal_2016}. Amyloid-$\beta$, which is strongly associated with widespread cortical tau pathology, can promote tau propagation and induce neuronal hyperexcitability~\cite{pooler_amyloid_2015,he_amyloid-_2018,busche_clusters_2008,zott_vicious_2019,costoya-sanchez_increased_2023,adams_distinct_2022}. Neuronal activity may therefore change the effective transmission of pathology through the structural network, altering the spatial course of disease and potentially driving the transition from localized to widespread pathology.

The established theory of spreading processes on networks offers a natural framework for addressing this limitation. Epidemic models have shown that heterogeneous activity, time-varying contact patterns, burstiness, and interactions between dynamical layers can alter spreading behavior~\cite{perra_activity_2012,pozzana_epidemic_2017,granell2013dynamical,mancastroppa_burstiness_2019,zhang_epidemic_2019,zino_analytical_2017,vazquez_impact_2007}. Recent models of neurodegeneration have similarly begun to couple neuronal activity to pathological protein progression~\cite{alexandersen_neuronal_2024,cabrera-alvarez_multiscale_2024,alexandersen_neuronal_2026,alexandersen_network_2026}. However, a general description of how neuronal activity interacts with brain-network structure to shape pathological spreading is still lacking.

Here, we address this gap by coupling a general node-activity process to susceptible--infected--susceptible dynamics on a fixed structural network. Activity modifies transmission from each source node, producing an activity-weighted spreading operator. The dominant eigenmode of this operator determines where a small pathological perturbation initially grows, while its leading eigenvalue determines the epidemic threshold separating decay from sustained growth~\cite{van_mieghem_virus_2009,gomez_discrete-time_2010,pastor-satorras_epidemic_2015}. This allows us to study how neuronal activity changes both the course of pathological spread and the conditions for disease onset.

A further challenge arises from the scale at which brain activity is measured. Activity varies among neurons within a brain region, but brain imaging averages their activity into a single regional measurement. This averaging hides fine-scale variation that may alter how pathology spreads between regions. To bridge these scales, we develop a multiscale description of how both measured regional activity and unresolved variation influence spreading across the brain.

We first derive perturbative approximations for the activity-dependent leading eigenvalue and dominant eigenmode. We find that the location of activity within the network matters. Activity concentrated in structurally influential regions most strongly shifts the epidemic threshold, whereas activity that couples different structural modes redirects where pathology initially grows. Stochastic simulations confirm these predictions across synthetic networks using a quiet--active--quiet process as a concrete neuron-like activity model. Furthermore, our multiscale analysis shows that fine-scale activity variation hidden by regional averaging can still influence spreading between regions, and that this influence can be approximated using regional estimates of activity variation rather than measurements of every neuron.

We next ask whether the theory's two central spectral quantities—the dominant eigenmode and epidemic threshold—predict where pathology progresses and whether it becomes widespread in the human brain. To test this, we apply the framework to longitudinal imaging data from the ADNI, HABS, and A4 studies~\cite{petersen_alzheimers_2010,dagley_harvard_2017,sperling_trial_2023}. We use a group structural connectome to define the spreading network and brain scans of regional glucose metabolism as a proxy for neuronal activity. Tau imaging measures how pathology progresses across brain regions over time. At the regional level, we test whether the activity-dependent dominant eigenmode predicts where tau pathology subsequently grows. Incorporating neuronal activity captures spatial patterns of disease progression beyond the structural eigenmode and established regional disease markers. At the individual level, lower activity-dependent epidemic thresholds are modestly associated with more widespread tau pathology, providing preliminary evidence that the threshold captures differences in susceptibility to widespread progression.

Together, these results show that the same spectral principles governing epidemics on networks also shape how Alzheimer's disease begins and spreads through the brain. Neuronal activity may therefore help us predict, and perhaps alter, the course of disease.

\section{Results}

The Results proceed in five steps. In Sec.~2.1, we derive the epidemic threshold and dominant eigenmode for activity-dependent spreading, establishing what determines whether and where pathological growth begins. In Sec.~2.2, we use perturbation theory to identify which activity patterns shift the threshold or redirect the dominant eigenmode, and validate them in stochastic simulations. In Sec.~2.3, we extend the theory across spatial scales to determine how regional mean activity and unresolved within-region variation influence spreading. In Sec.~2.4, we test whether the predicted dominant eigenmode explains where tau pathology progresses in longitudinal brain imaging. Finally, in Sec.~2.5, we test whether the activity-dependent epidemic threshold is associated with how broadly tau spreads across individuals.

\subsection{An activity-dependent epidemic threshold and dominant eigenmode}

To determine how neuronal activity influences pathological spreading, we first derive the epidemic threshold, which determines whether small pathological seeds grow, and the dominant invasion mode, which determines where growth begins. We represent pathological spreading on a graph using susceptible--infected--susceptible (SIS) dynamics, in which infection propagates along edges from infected to susceptible nodes, and infected nodes can clear the infection and become susceptible again. We couple this spreading process to a general node-activity process in which each node is either quiet or active. Both processes evolve on a weighted network with nonnegative, irreducible adjacency matrix
\(W\in\mathbb{R}_{\geq 0}^{N\times N}\), where \(W_{ij}\) denotes the
connection weight from source node \(j\) to target node \(i\). Let
\(p^I(t),p^A(t)\in[0,1]^N\) denote the corresponding node-wise marginal
probabilities of infection and activity. Motivated by experimental evidence that neuronal activity promotes tau release and propagation~\cite{pooler_physiological_2013,yamada_neuronal_2014,wu_neuronal_2016}, and following earlier activity-dependent spreading models~\cite{alexandersen_neuronal_2024,alexandersen_neuronal_2026}, we let activity at source node \(j\) rescale transmission along its outgoing edges by the factor \(1+\delta p_j^A(t)\), where $-1 < \delta < \infty$. Thus, when \(\delta>0\), more active nodes transmit infection more strongly. Following the microscopic Markov chain
approach~\cite{gomez_discrete-time_2010,gomez_nonperturbative_2011,
chakrabarti_epidemic_2008}, the infection probabilities evolve approximately as
\begin{equation*}
\begin{aligned}
p^I(t+1)
=
p^I(t)
+
\varepsilon
\Big[
\bigl(\mathbf 1-p^I(t)\bigr)
\odot
\beta W
\operatorname{diag}\!\bigl(\mathbf 1+\delta p^A(t)\bigr)
p^I(t)
-
\zeta p^I(t)
\Big],
\end{aligned}
\label{eq:general_activity_sis}
\end{equation*}
where \(\odot\) denotes element-wise multiplication, \(\beta\geq0\) is the
transmission rate, \(\zeta \geq 0\) is the clearance rate, and
\(\varepsilon>0\) scales the SIS update relative to the activity
update.

We leave the activity dynamics unspecified
and write them as
\begin{equation*}
p^A(t+1)
=
\mathcal G\bigl(p^A(t),p^I(t)\bigr).
\label{eq:general_activity_map}
\end{equation*}
Assume that the disease-free activity subsystem has a stable fixed
point $a=\mathcal G(a,\mathbf 0)$, satisfying
$
\rho(J_A)<1,
$
where \(\mathbf 0\in\mathbb R^N\) denotes the all-zero vector and
\(\rho(\cdot)\) denotes the spectral radius. 
The full coupled system then
admits the disease-free fixed point
$
(p^I,p^A)=(\mathbf 0,a).$
Linearizing the coupled dynamics around this state gives the
block-triangular Jacobian
\begin{equation}
J_{\mathrm{DF}}
=
\begin{pmatrix}
J_I & 0_{N}\\
C & J_A
\end{pmatrix},
\label{eq:block_triangular_dfe_jacobian}
\end{equation}
where
\begin{equation*}
J_I
=
(1-\varepsilon\zeta)I
+
\varepsilon\beta W A,
\qquad
J_A
=
\left.
\frac{\partial \mathcal G}{\partial p^A}
\right|_{(p^A,p^I)=(a,\mathbf 0)},
\qquad
C
=
\left.
\frac{\partial \mathcal G}{\partial p^I}
\right|_{(p^A,p^I)=(a,\mathbf 0)},
\label{eq:infection_jacobian}
\end{equation*}
and
\[
A
=
\operatorname{diag}(\mathbf 1+\delta a),
\]
where, for $x \in \mathbb{R}^N$, $\operatorname{diag}(x)$ is the $N \times N$ diagonal matrix with entries $\operatorname{diag}(x)_{ii} = x_i$, \(I\) is the \(N\times N\) identity matrix, \(0_{N}\) is the \(N\times N\) zero matrix, and
\(\mathbf 1\in\mathbb R^N\) is the all-one vector.

The block structure isolates the stability condition that determines whether pathology can invade. The upper-right block in Eq.~\eqref{eq:block_triangular_dfe_jacobian} vanishes because activity modifies transmission only in the presence of infection. Consequently, a perturbation of \(p^A\) enters the infection dynamics multiplied by \(p^I\) and is therefore second order at the disease-free state. Because the Jacobian is block triangular, its eigenvalues are the union of those of the infection and activity blocks. The activity block is stable by assumption, so a loss of disease-free stability can occur only through the infection block \(J_I\), which therefore determines the epidemic threshold.
Because \(A\) is a strictly positive diagonal matrix and \(W\) is
nonnegative and irreducible, \(WA\) is also nonnegative and
irreducible. Therefore,
\(J_I\) is likewise
nonnegative and irreducible given $1-\varepsilon \zeta \geq 0$,
with Perron--Frobenius eigenvalue
\begin{equation*}
\rho(J_I)
=
1-\varepsilon\zeta
+
\varepsilon\beta\rho(WA).
\end{equation*}
The disease-free state is therefore stable when
$
\beta\rho(WA)<\zeta,
$
giving the activity-dependent epidemic threshold
\begin{equation*}
\beta_c(a)
=
\frac{\zeta}{
\rho\!\left(
W A
\right)
}.
\label{eq:activity_dependent_threshold}
\end{equation*}

The same activity-weighted spreading operator also determines the
spatial pattern of local invasion. Let \(r_0(a)\) denote its right
Perron mode, satisfying
$
WA r_0(a)
=
\rho(W A) r_0(a).
$
Because \(J_I\) is an affine function of \(WA\), the two matrices have
the same right eigenvectors, and
\begin{equation*}
J_I r_0(a)
=
\left[
1-\varepsilon\zeta
+
\varepsilon\beta\rho(WA)
\right]r_0(a).
\end{equation*}
At $\beta=\beta_c(a)$, the Perron--Frobenius eigenvalue of $J_I$ reaches unity. Provided $1-\varepsilon\zeta>0$, $J_I$ is a primitive matrix, so this eigenvalue is uniquely dominant and all other eigenvalues lie strictly inside the unit circle. Immediately above the threshold, a generic small infection perturbation therefore grows predominantly along $r_0(a)$ during the initial linear phase. Thus, $r_0(a)$ defines the relative regional pattern of early infection
growth.

These results show two distinct ways in which neuronal activity can shape pathological spreading. By changing the leading eigenvalue of the activity-weighted network, activity can make invasion easier or harder. By changing its dominant mode, it can alter where growth first concentrates. At this early stage, the activity process enters only through its stable disease-free pattern \(a\). This leads to the central question of the next section: which spatial patterns of activity most strongly shift the threshold, and which redirect the initial pattern of spread?

\subsection{Spectral perturbations of the activity-dependent threshold and dominant eigenmode}
\label{sec:results_threshold_verification}

We now determine how activity shifts the epidemic threshold and redirects the dominant invasion mode. Expanding in activity-dependent transmission identifies the patterns driving each change. The
epidemic threshold and Perron mode described above are determined by the spectral properties
of the effective spreading matrix
\[
M(\delta)
=
WA
=
W(I+\delta D_a),
\qquad
D_a=\operatorname{diag}(a).
\]
Its spectral radius controls
whether a small infection perturbation grows or decays, while its
Perron mode gives the spatial
direction that first becomes unstable when the epidemic threshold is
crossed. 

We consider a symmetric, nonnegative, and irreducible adjacency matrix
\(W\), with orthonormal eigenpairs
\[
Wq_k=\lambda_kq_k,
\qquad
q_k^\top q_\ell=\delta_{k\ell},
\qquad
\lambda_0=\rho(W),
\]
where $\delta_{kl}$ is the Kronecker delta.
Let \(r_0(\delta)\) denote the Perron mode of
\(M(\delta)\), normalized such that
$
q_0^\top r_0(\delta)=1.
$
We perturb both the leading eigenvalue and Perron mode in the regime of weak activity-to-spreading coupling
\(|\delta|\ll1\). Defining
$
d_{k\ell}
=
q_k^\top D_aq_\ell
$
and
$
\Delta_k
=
\lambda_0-\lambda_k,
$
we obtain
\begin{equation}
\label{eq:symmetric_spectral_radius}
\rho(M(\delta))
=
\lambda_0
\left[
1
+
\delta d_{00}
+
\delta^2
\sum_{k\geq1}
\frac{\lambda_k}{\Delta_k}
d_{k0}^2
\right]
+
O(\delta^3),
\end{equation}
and
\begin{equation}
\label{eq:perron_vector_expansion}
r_0(\delta)
=
q_0
+
\delta r^{(1)}
+
\delta^2r^{(2)}
+
O(\delta^3),
\end{equation}
where
\begin{equation}
\label{eq:perron_vector_first_order}
r^{(1)}
=
\sum_{k\geq1}
\frac{\lambda_k}{\Delta_k}
d_{k0}q_k,
\end{equation}
and
\begin{equation}
\label{eq:perron_vector_second_order}
r^{(2)}
=
\sum_{k\geq1}
\frac{\lambda_k}{\Delta_k}
\left[
\sum_{\ell\geq1}
\frac{\lambda_\ell}{\Delta_\ell}
d_{k\ell}d_{\ell0}
-
\frac{\lambda_0}{\Delta_k}
d_{00}d_{k0}
\right]q_k.
\end{equation}
The derivation and normalization details are given in Supporting
Information Sec.~A.

The quantity
$
d_{k\ell}
=
q_k^\top D_aq_\ell
$
is the signed component of the activity-weighted vector \(D_aq_\ell\)
along structural mode \(q_k\). In particular, \(d_{k0}\) measures how
activity mixes the unperturbed Perron mode with mode \(q_k\). This
mixing changes the dominant eigenmode at first order and feeds back
into the leading eigenvalue at second order. The second-order
mode term additionally accounts for two-step mixing through
intermediate structural modes. Both the threshold and spatial mode are
therefore especially sensitive to activity patterns that couple the
Perron mode to structural modes separated from it by a small spectral
gap.

If
activity is spatially uniform, \(a=a_0\mathbf 1\), then
$
M(\delta)
=
(1+\delta a_0)W,
$
rescaling the spectral radius, but leaving the Perron mode. Similarly, for a rank-one network
$
W
=
\lambda_0q_0q_0^\top,
$
we have
\begin{equation}
\rho(M(\delta))
=
\lambda_0
\left(
1+\delta q_0^\top D_aq_0
\right),
\qquad
r_0(\delta)=q_0
\label{eq:rank1-spectrum}
\end{equation}
for an arbitrary activity pattern. The homogeneous all-to-all network
with self-loops is the special case \(q_0=N^{-1/2}\mathbf 1\), for which the spectral radius
$
\rho(M(\delta))
=
1+\delta \sum_{=1}^N a_i/N
$
depends simply on the average network activity.
Thus, activity can shift the epidemic threshold without changing the
dominant eigenmode. A nontrivial redistribution of the mode
requires both heterogeneous activity and nonzero subdominant
structural modes.

\paragraph{Validation against stochastic simulations.}
We next tested whether the spectral approximations accurately capture the epidemic threshold and dominant mode of a coupled stochastic activity--spreading process. To do so, we instantiated the general
activity process with the stochastic QAQ dynamics and coupled it to SIS
spreading as defined in Sec.~\ref{sec:method_MMCA}. The QAQ process
provides a concrete neuron-like mechanism for generating the
disease-free activity pattern \(a\) and stochastic activity
trajectories.
Epidemic thresholds were estimated from quasi-stationary simulations of
the full SISQAQ discrete time Markov chain (DTMC) using the procedure described in
Sec.~\ref{sec:method_numerical}.

\begin{figure}[t]
    \centering
    \includegraphics[width=\linewidth]{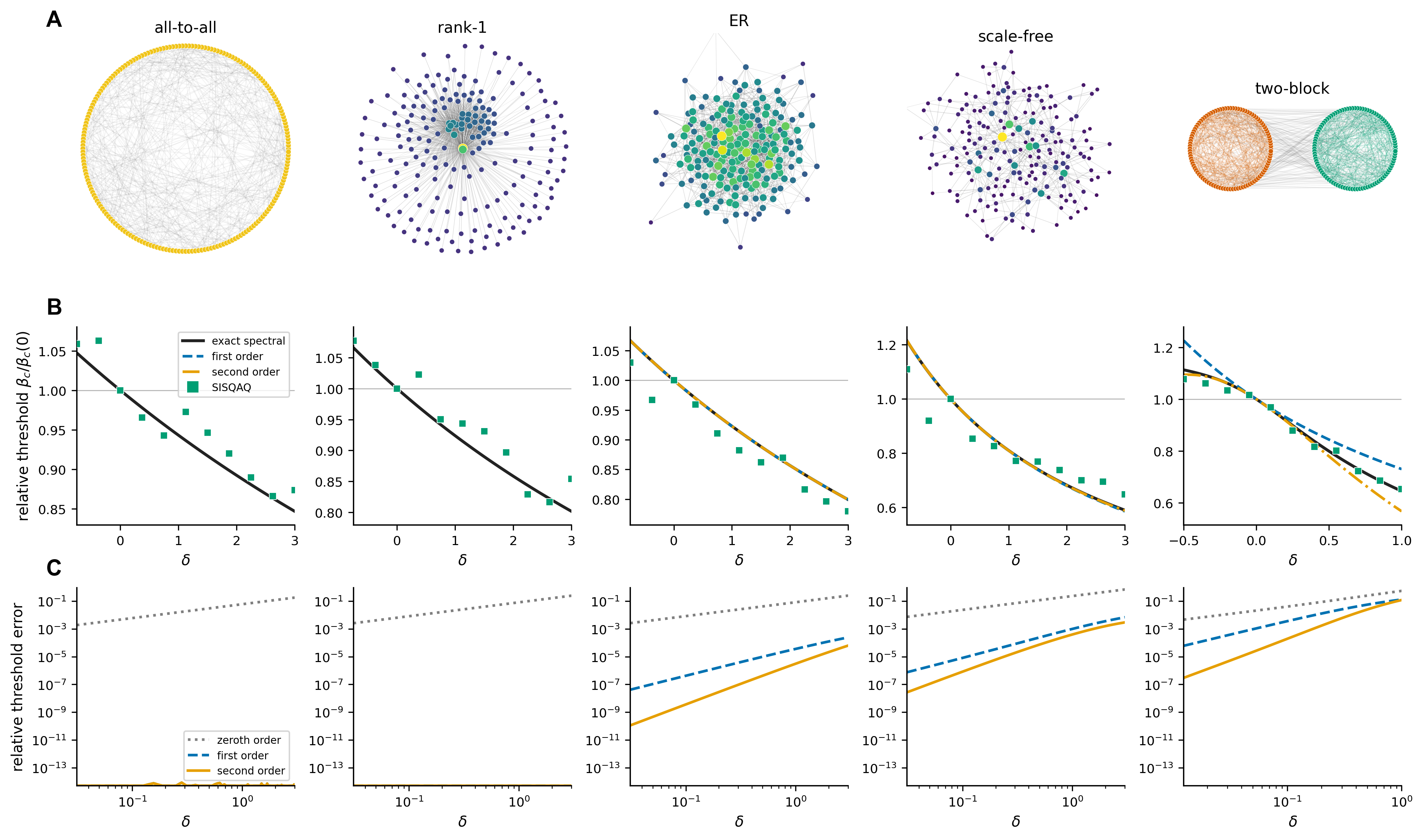}
    \caption{\textbf{Threshold predictions across synthetic network families.}
    (\textbf{A}) Schematics of the five networks used in the comparison. Node color and size encode weighted degree in the rank-one, Erd\H{o}s--R\'enyi, and scale-free networks. Nodes in the all-to-all network are equivalent, while colors in the two-block network indicate block membership. (\textbf{B}) Relative epidemic threshold $\beta_c(\delta)/\beta_c(0)$ as a function of the activity--infection coupling $\delta$. Lines show the exact spectral threshold and its first- and second-order approximations; squares denote threshold estimates from quasi-stationary SISQAQ simulations. (\textbf{C}) Relative errors of the zeroth-, first-, and second-order threshold approximations. The networks were not matched in average degree. All had $N=200$ and were rescaled so that $\rho(W)=1$. Baseline QAQ parameters were $\gamma=0.34$, $I_c^0=0.73$, $\sigma_\varepsilon=0.33$, and $\alpha=1.05$; the two-block network used $\gamma_1=0.175$ and $\gamma_2=0.90$ to produce heterogeneous baseline activity.}
    \label{fig:network_thresholds}
\end{figure}

As shown in Fig.~\ref{fig:network_thresholds}, the exact spectral
thresholds closely tracked the simulation-based thresholds across five
synthetic network families. For the all-to-all and rank-one networks,
the perturbation series is exact as expected from Eq.~\eqref{eq:rank1-spectrum}. For the Erd\H{o}s--R\'enyi and
scale-free networks, the first- and second-order approximations both
followed the exact threshold closely, with the second-order approximation
giving the smaller error. The two-block network provides a case in
which the second-order correction is more important. For two equal
blocks of size \(N/2\), with within-block weight \(w_{\mathrm{in}}\),
between-block weight \(w_{\mathrm{out}}\), and no self-loops, the two
leading eigenvalues are
$
\lambda_0
=
\left(\frac{N}{2}-1\right)w_{\mathrm{in}}
+
\frac{N}{2}w_{\mathrm{out}},
$
and
$
\lambda_1
=
\left(\frac{N}{2}-1\right)w_{\mathrm{in}}
-
\frac{N}{2}w_{\mathrm{out}}.
$
Their spectral gap is
$
\lambda_0-\lambda_1
=
Nw_{\mathrm{out}},
$
which vanishes as \(w_{\mathrm{out}}\to0\). The small spectral gap
amplifies the second-order correction when the activity pattern has a
nonzero projection onto the block-contrast mode. Accordingly, within
\(\lvert\delta\rvert\leq0.5\), the second-order threshold error in the
two-block network was reduced more than fivefold compared with first
order (RMSE \(0.0037\) versus \(0.0203\)). Thus, the leading-order
approximation is sufficient when the dominant spreading mode is well-separated, whereas higher-order corrections become important in
modular networks with near-degenerate modes and heterogeneous activity.

\begin{figure}[t]
    \centering
    \includegraphics[width=\linewidth]{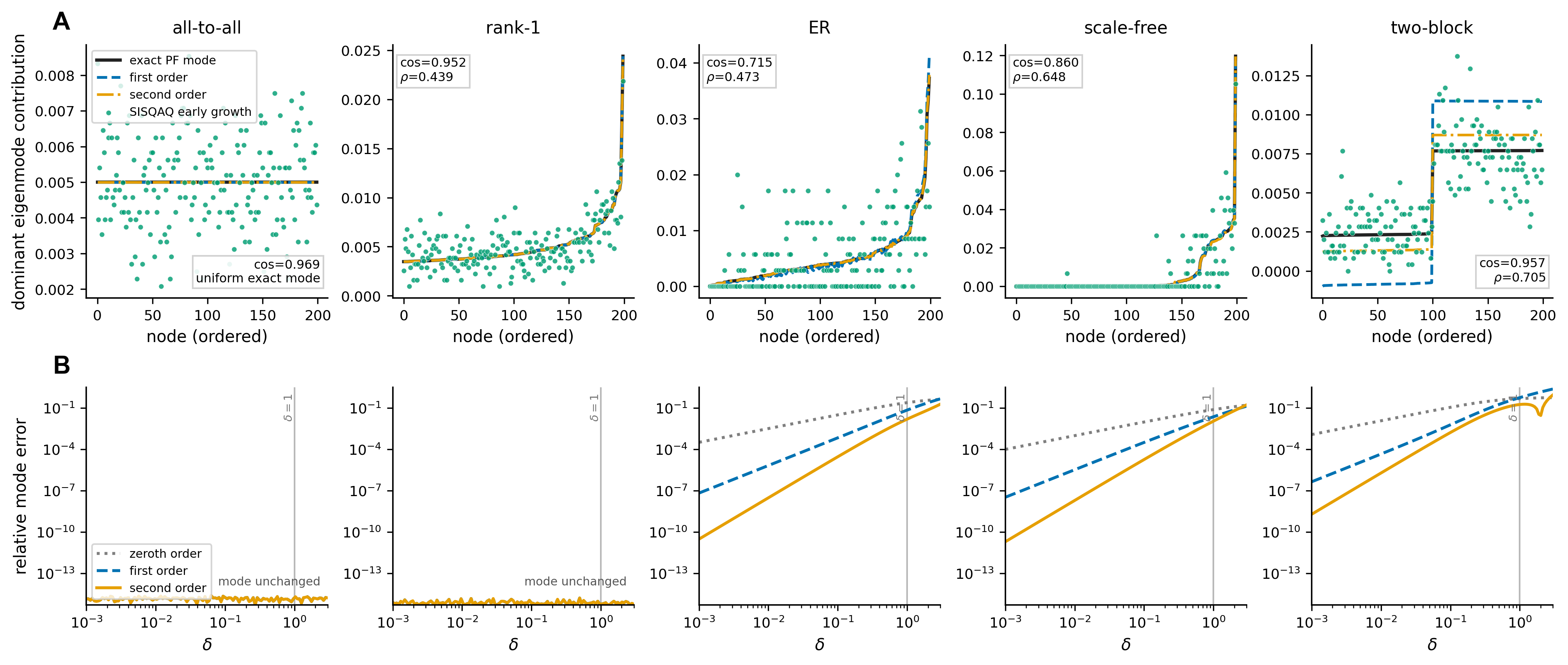}
    \caption{\textbf{Validation of the activity-dependent dominant eigenmode.}
    (\textbf{A}) Comparison of the exact normalized Perron mode components, first- and second-order approximations, and mean early-growth estimates from stochastic SISQAQ simulations across five network families. Nodes are ordered by the exact Perron mode. (\textbf{B}) Relative errors of the zeroth-, first-, and second-order Perron modes as $\delta$ is varied. Vertical lines mark $\delta=1$, as used for the stochastic profiles. SISQAQ profiles were averaged over 384 runs at $\beta=1.08\beta_c$, with infection initially seeded in 1\% of nodes.}
    \label{fig:perron_mode_validation}
\end{figure}

We then tested whether the Perron mode predicts the spatial
profile of incipient stochastic spreading. For each network, we
compared the exact Perron mode and its first- and second-order
approximations with the node-wise infection profile during early
SISQAQ growth (Fig.~\ref{fig:perron_mode_validation}). The stochastic
profiles were aligned with the exact modes, which are well-approximated by the first- and second-order approximations. The
all-to-all and rank-one Perron modes were unchanged by activity
as expected from Eq.~\eqref{eq:rank1-spectrum}. 

\begin{figure}[!ht]
    \centering
    \includegraphics[width=\linewidth]{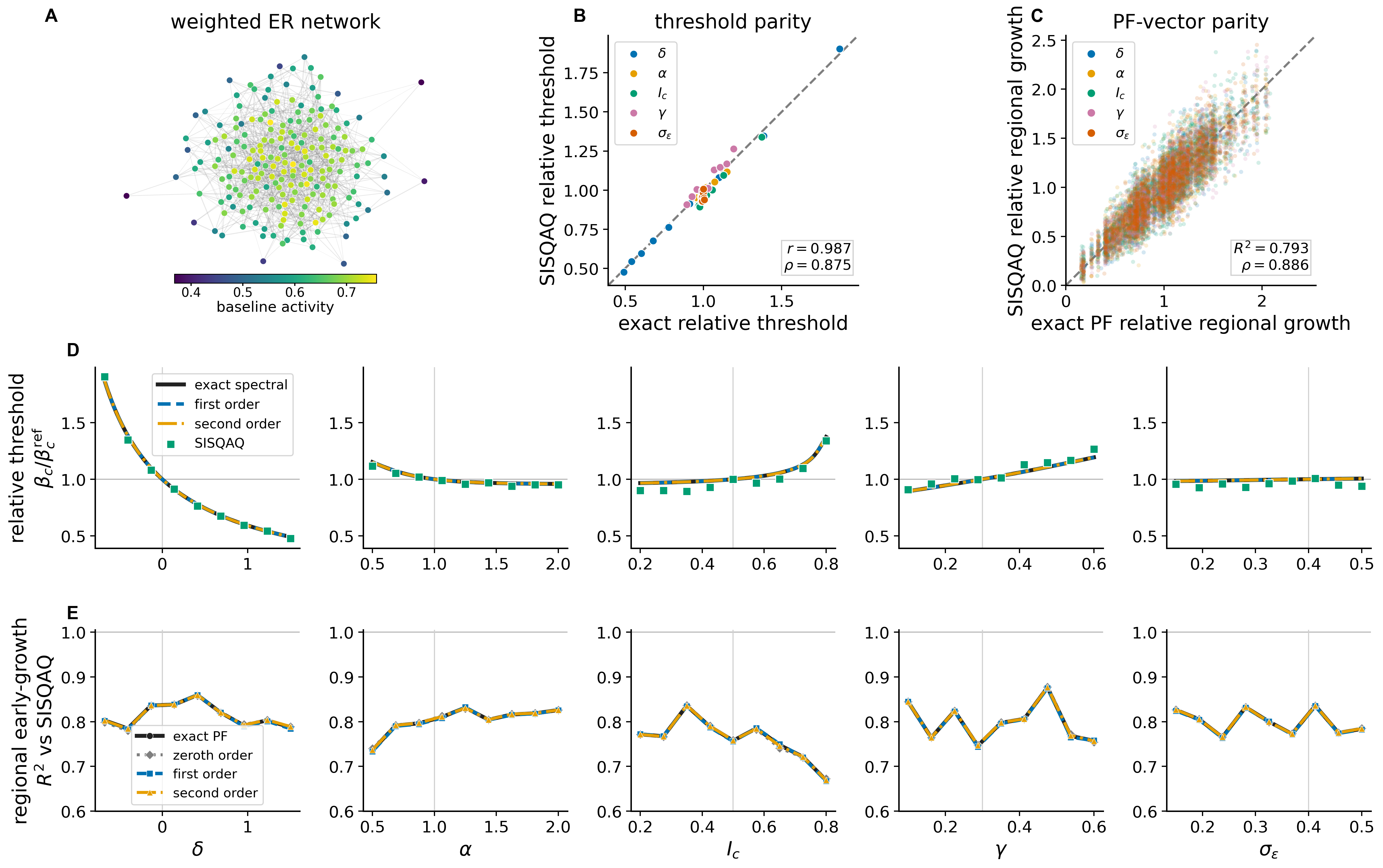}
    \caption{\textbf{Threshold and dominant eigenmode validation across QAQ parameter sweeps on a heterogeneous network.}
    (\textbf{A}) Weighted Erd\H{o}s--R\'enyi network with \(N=200\), connection probability \(p=0.05\), and \(\rho(W)=1\). (\textbf{B}) Parity of exact and SISQAQ relative thresholds across sweeps in \(\delta\), \(\alpha\), \(I_c\), \(\gamma\), and \(\sigma_\varepsilon\). (\textbf{C}) Parity of the exact Perron mode and SISQAQ regional early growth across the same parameter sweeps. Colors identify the varied parameter, and dashed diagonals denote equality; profiles in panel (\textbf{C}) are normalized by their network mean. (\textbf{D}) Relative epidemic thresholds for the five parameter sweeps. Black, blue dashed, and orange dash-dotted curves show the exact, first-order, and second-order spectral predictions, respectively; green squares show quasi-stationary SISQAQ estimates. (\textbf{E}) Node-wise \(R^2\) between each predicted Perron mode and the corresponding SISQAQ early-growth profile. Black, gray dotted, blue dashed, and orange dash-dotted curves show the exact, zeroth-, first-, and second-order modes, respectively.}
    \label{fig:er_threshold_perron_parameter_sweeps}
\end{figure}

Finally, we tested the spectral perturbations across a broader range of QAQ
activity regimes. We varied
\(\delta\), \(\alpha\), \(I_c\), \(\gamma\), and
\(\sigma_\varepsilon\) one at a time on a fixed weighted
Erd\H{o}s--R\'enyi network
(Fig.~\ref{fig:er_threshold_perron_parameter_sweeps}; see Methods for parameter descriptions). Varying
\(\delta\) changes the coupling from activity to transmission, whereas
varying the QAQ parameters changes the heterogeneous disease-free
activity pattern entering
\(D_a=\operatorname{diag}(a)\).
Across the 45 parameter settings, the simulation-based epidemic
thresholds closely tracked the exact thresholds (Pearson \(r=0.987\) and
Spearman \(\rho=0.875\)). As individual QAQ parameters varied, the
first- and second-order spectral thresholds also closely followed the
exact and simulation-based thresholds.

The Perron mode approximations showed similarly consistent behavior (Fig.~\ref{fig:er_threshold_perron_parameter_sweeps}). Across parameter settings and nodes, the exact Perron mode closely tracked SISQAQ early growth
(\(R^2=0.793\) and Spearman \(\rho=0.886\)). At individual parameter
settings, the exact Perron mode explained a median \(79.7\%\) of the
regional variation in SISQAQ early growth, with a range of
\(66.9\%\)--\(87.6\%\). 

Having validated the spectral approximations, we next extend them across spatial scales to distinguish the effects of regional mean activity from variation hidden within regions.

\subsection{Multiscale decomposition of the activity-dependent threshold and dominant eigenmode}
\label{sec:results_multiscale_spectral}

Brain imaging measures activity at the level of regions, averaging over
the neurons within each region. This coarse-graining hides fine-scale
variation that may nevertheless alter how pathology spreads between
regions. To determine how both regional mean activity and unresolved
within-region variation shape spreading, we use a block-network
description in which each block represents a brain region and the nodes
within it resolve activity at a finer spatial scale.

Let \(W\in\mathbb R^{n\times n}_{\geq0}\) be a symmetric, nonnegative,
irreducible, and equitable block matrix with \(B\) blocks of equal size
\(M\), so that \(n=BM\). We decompose the disease-free activity pattern
within block \(b\) as
\[
a_{bi}
=
\bar a_b+\theta_{bi},
\qquad
\sum_{i=1}^M\theta_{bi}=0,
\]
or equivalently,
\[
D_a
=
D_{\bar a}+D_\theta.
\]
Here, $D_{\bar{a}} \in \mathbb{R}^{n \times n}$ is the diagonal matrix containing block-averages repeated in each block $D_{\bar{a},ii} = a_b$ for $i$ in block $b$, and $D_\theta \in \mathbb{R}^{n \times n}$ is the diagonal matrix containing the node-specific deviation from the block mean $D_{\theta,ii} = \theta_{bj}$ where node $i$ is the $j$-th node in block $b$.

The equitable partition defines the block-constant subspace
\[
U
=
\left\{
x\in\mathbb R^n:
x_{bi}=x_b
\text{ for all }i
\right\},
\]
and its orthogonal complement
\[
U^\perp
=
\left\{
x\in\mathbb R^n:
\sum_{i=1}^M x_{bi}=0
\text{ for every }b
\right\}.
\]
Because \(W\) is symmetric and equitable, both \(U\) and \(U^\perp\)
are invariant under \(W\), and its eigenvectors may be chosen to lie in
one of these two subspaces. We refer to eigenvectors in \(U\) as block
modes and eigenvectors in \(U^\perp\) as internal
modes~\cite{brouwer_spectra_2011}.

Let \(q_0\) denote the normalized Perron mode of \(W\). Then
\(q_0\in U\). Block-average activity preserves both subspaces,
$
D_{\bar a}U\subseteq U$ and
$D_{\bar a}U^\perp\subseteq U^\perp$.
Whereas, centered deviations satisfy
$
D_\theta U\subseteq U^\perp.
$
These
selection rules together with the orthogonality of block and internal eigenvectors determine which perturbative contributions remain
visible after coarse-graining to the quotient network, whose nodes represent blocks and whose connection weights summarize connectivity between blocks. Details are
given in Supporting Information Sec.~B.

Applying the decomposition to the leading-eigenvalue expansion Eq.~\eqref{eq:activity_dependent_threshold} gives
\begin{equation}
\label{eq:block_spectral_radius}
\begin{aligned}
\rho(M(\delta))
=
\lambda_0
\Bigg[
1
&+
\delta q_0^\top D_{\bar a}q_0
\\
&+
\delta^2
\sum_{\substack{k\geq1\\q_k\in U}}
\frac{\lambda_k}{\lambda_0-\lambda_k}
\left(q_k^\top D_{\bar a}q_0\right)^2
\\
&+
\delta^2
\sum_{\substack{\alpha\\q_\alpha\in U^\perp}}
\frac{\lambda_\alpha}{\lambda_0-\lambda_\alpha}
\left(q_\alpha^\top D_\theta q_0\right)^2
\Bigg]
+
O(\delta^3).
\end{aligned}
\end{equation}
Block-average activity therefore affects the epidemic threshold at
first and second order. Within-block deviations have no first-order
contribution because
$
q_0^\top D_\theta q_0=0,
$
and first affect the threshold at second order through their overlap
with internal modes.

The leading eigenvector Eq.~\eqref{eq:perron_vector_expansion} separates similarly. Decomposing the first-order term Eq.~\eqref{eq:perron_vector_first_order} gives
\begin{equation}
\label{eq:block_perron_first_order}
\begin{aligned}
r^{(1)}
=
&
\sum_{\substack{k\geq1\\q_k\in U}}
\frac{\lambda_k}{\lambda_0-\lambda_k}
\left(q_k^\top D_{\bar a}q_0\right)q_k
\\
&+
\sum_{\substack{\alpha\\q_\alpha\in U^\perp}}
\frac{\lambda_\alpha}{\lambda_0-\lambda_\alpha}
\left(q_\alpha^\top D_\theta q_0\right)q_\alpha.
\end{aligned}
\end{equation}
The first sum lies in \(U\) and changes the dominant eigenmode of the coarse-grained quotient matrix. The second term, however, lies in \(U^\perp\) which vanishes under coarse-graining. Thus, even though activity deviations $D_\theta$ alter the fine-scale spatial mode at first
order, this effect disappears when coarse-graining the network to the block level.

Applying the decomposition to the second-order term Eq.~\eqref{eq:perron_vector_second_order}, similarly results in two sums where each is either in $U$ or $U^\perp$. The latter vanishes under coarse-graining and is left out. 
Defining
\[
b_{k\ell}
=
q_k^\top D_{\bar a}q_\ell,
\qquad
t_{k\ell}
=
q_k^\top D_\theta q_\ell,
\]
the block-level second-order correction is
\begin{equation}
\label{eq:block_perron_second_order}
\begin{aligned}
\Pi_Ur^{(2)}
=
\sum_{\substack{k\geq1\\q_k\in U}}
\frac{\lambda_k}{\lambda_0-\lambda_k}
\Bigg[
&
\sum_{\substack{\ell\geq1\\q_\ell\in U}}
\frac{\lambda_\ell}{\lambda_0-\lambda_\ell}
b_{k\ell}b_{\ell0}
-
\frac{\lambda_0}{\lambda_0-\lambda_k}
b_{00}b_{k0}
\\
&+
\sum_{\substack{\alpha\\q_\alpha\in U^\perp}}
\frac{\lambda_\alpha}{\lambda_0-\lambda_\alpha}
t_{k\alpha}t_{\alpha0}
\Bigg]q_k.
\end{aligned}
\end{equation}
The first line is the second-order mean-activity response. The second
line is the part of the internal deviation response that is visible at the quotient level. 

The latter term above shows that internal deviations in activity indeed do have an influence on the block-level dynamics. This is of interest to us, because internal deviations, which relate to internal regional variations in the structural connectome, are usually ignored. We now show that we may not need precise measurements of each individual neuron, but instead rely on population level measurements of activity variance to account for the second-order internal deviation term in Eq.~\eqref{eq:block_perron_second_order}.

Neuroimaging resolves regional quantities but not the internal
activity pattern or internal network structure within each region. We
therefore parameterize the unresolved activity within block \(b\) as
\[
a_{bi}
=
u_b
+
\sigma\sqrt{Mv_b}\,s_{bi},
\]
where \(u_b\geq0\) is the regional mean activity, \(v_b\geq0\) controls
the magnitude of within-region heterogeneity, and
\(s_b\in\mathbb R^M\) is a centered, unit-norm internal contrast,
\[
\sum_i s_{bi}=0,
\qquad
\sum_i s_{bi}^2=1.
\]
This ensures that
\[
\frac{1}{M}\sum_i a_{bi}=u_b,
\qquad
\frac{1}{M}\sum_i(a_{bi}-u_b)^2
=
\sigma^2v_b.
\]

For a block mode \(q_k\in U\), define its quotient coordinates
\(\widetilde q_k\in\mathbb R^B\) by
$
q_{k,bi}
=
\widetilde q_{k,b} / \sqrt M.
$
To represent the unresolved internal contribution without specifying
the full within-region network, we assume one localized internal mode
per region. Let \(W_{cb}\) denote the block mapping source block \(b\)
to target block \(c\), and assume
$
W_{bb}s_b=\mu_bs_b,
$
and
$
W_{cb}s_b=0
$
for
$
c\neq b.
$
Embedding this block-specific mode into the full network by setting it to zero outside block $b$ gives
$
\widehat s_b
=
[0,\ldots,0,s_b,0,\ldots,0]^\top
$
which is an internal eigenvector of the full network,
\[
W\widehat s_b=\mu_b\widehat s_b.
\]
This localization assumption is stronger than equitability alone and
is satisfied, for example, when between-block connectivity acts only
on block means.

Define
\[
\chi_b
=
\frac{\mu_b}{\lambda_0-\mu_b},
\]
and, for block modes,
$
m_{k\ell}(u)
=
\widetilde q_k^\top
\operatorname{diag}(u)
\widetilde q_\ell.
$
Under this parameterization, the leading eigenvalue Eq.~\eqref{eq:block_spectral_radius} becomes
\begin{equation}
\label{eq:brain_epidemic_threshold}
\begin{aligned}
\rho(M(\delta))
=
\lambda_0
\Bigg[
1
&+
\delta m_{00}(u)
\\
&+
\delta^2
\sum_{k\in\mathcal B}
\frac{\lambda_k}{\lambda_0-\lambda_k}
m_{k0}(u)^2
\\
&+
\delta^2\sigma^2
\sum_b
\widetilde q_{0,b}^2
v_b\chi_b
\Bigg]
+
O(\delta^3),
\end{aligned}
\end{equation}
where \(\mathcal B\) denotes the non-Perron block modes. The final term
is the contribution of unresolved activity heterogeneity to the
epidemic threshold.

As for the Perron eigenmode perturbation, the first-order internal contribution is
\begin{equation*}
\label{eq:internal_perron_first_order}
r_\theta^{(1)}
=
\sigma
\sum_b
\widetilde q_{0,b}\sqrt{v_b}\,
\chi_b\widehat s_b,
\end{equation*}
which vanishes when coarse-graining to the quotient matrix, hence the first-order Perron mode perturbation on the quotient level does not depend on $D_\theta$ only $D_{\bar a}$.
To express the quotient-scale mode compactly, define
\begin{equation}
\label{eq:quotient_response_operator}
R
=
\sum_{k\in\mathcal B}
\frac{\lambda_k}{\lambda_0-\lambda_k}
\widetilde q_k\widetilde q_k^\top
\operatorname{diag}(\widetilde q_0).
\end{equation}
The first-order term is then
\[
\widetilde r_u^{(1)}
=
Ru.
\]
The quotient-scale second-order perturbation generated by mean activity $D_{\bar a}$ is 
\begin{equation}
\label{eq:second_order_mean_perron_response}
\begin{aligned}
\widetilde r_{uu}^{(2)}
=
\sum_{k\in\mathcal B}
\frac{\lambda_k}{\lambda_0-\lambda_k}
\Bigg[
&
\sum_{\ell\in\mathcal B}
\frac{\lambda_\ell}{\lambda_0-\lambda_\ell}
m_{k\ell}(u)m_{\ell0}(u)
\\
&-
\frac{\lambda_0}{\lambda_0-\lambda_k}
m_{00}(u)m_{k0}(u)
\Bigg]\widetilde q_k,
\end{aligned}
\end{equation}
while the second-order term generated by internal deviations is
\begin{equation}
\label{eq:second_order_internal_perron_response}
\widetilde r_{\theta\theta}^{(2)}
=
\sigma^2R(v\odot\chi).
\end{equation}
Combining these terms gives the quotient-scale dominant eigenmode,
\begin{equation}
\label{eq:brain_perron_mode}
\widetilde r_0(\delta)
=
\widetilde q_0
+
\delta Ru
+
\delta^2
\left[
\widetilde r_{uu}^{(2)}
+
\sigma^2R(v\odot\chi)
\right]
+
O(\delta^3).
\end{equation}
At the
regional (quotient) scale, mean activity contributes at first and second order,
whereas centered within-region deviations first become visible at
second order. Details of the derivation of Eqs.~\eqref{eq:brain_epidemic_threshold} and~\eqref{eq:brain_perron_mode} are given in Supporting Information Sec.~B.

\begin{figure}[t]
    \centering
    \includegraphics[width=\linewidth]{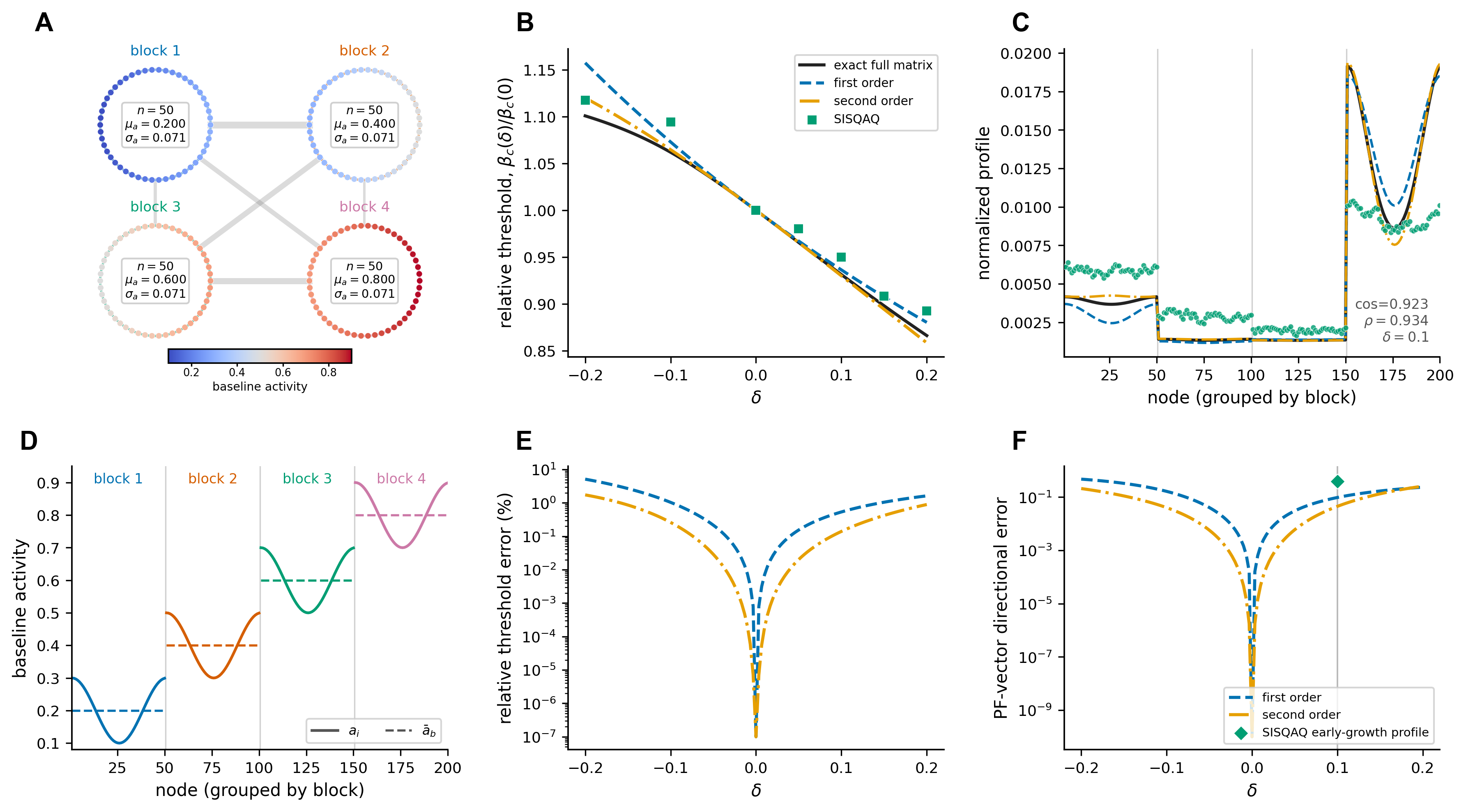}
    \caption{\textbf{Numerical validation of the multiscale threshold and dominant eigenmode.}
    (\textbf{A}) Synthetic four-block weighted network with a slowly decaying centered internal mode; node colors indicate baseline activity and each block contains 50 nodes. (\textbf{B}) Exact, first-order, and second-order relative epidemic thresholds compared with quasi-stationary SISQAQ estimates. (\textbf{C}) Exact and approximate Perron modes compared with the SISQAQ early-growth infection profile at $\delta=0.1$. (\textbf{D}) Node-level baseline activity and the corresponding block averages. (\textbf{E}) Relative errors of the first- and second-order threshold approximations. (\textbf{F}) Directional errors of the first- and second-order Perron-mode approximations; the diamond denotes the SISQAQ early-growth profile.}
    \label{fig:multiscale_decomposition_validation}
\end{figure}

We tested the full blockwise decomposition on a four-block equitable
network with distinct block-average activities, centered within-block
deviations, and a slowly decaying internal mode separated from the
Perron root by a spectral gap of \(0.029\)
(Fig.~\ref{fig:multiscale_decomposition_validation}). At \(\delta=0.1\), including the
second-order terms reduced the error relative to the exact threshold
from \(0.536\%\) to \(0.140\%\), and reduced the Perron-mode directional error from
\(0.0970\) to \(0.0451\). Quasi-stationary SISQAQ simulations reproduced
the activity-dependent change in the threshold, while the early-growth
infection profile agreed with the exact Perron mode (cosine similarity
\(=0.923\), Spearman \(\rho=0.934\)).

\subsection{The activity-dependent dominant eigenmode predicts future tau accumulation}

We next ask whether the spectral changes derived above are reflected in the
spatial progression of Alzheimer's disease in the human brain. If neuronal
activity redirects pathological spreading, then the activity-dependent
dominant mode should identify where tau accumulates over time beyond the
unperturbed structural mode and established disease markers. Longitudinal
tau-PET provides a regional empirical measure of this accumulation.

For this application, we use a group-averaged structural connectome as
the quotient network. We estimate \(u_b\) using the population mean
FDG-PET signal in region \(b\), and use the population variance of
regional FDG-PET as a surrogate for \(v_b\).

\paragraph{Parameter-free eigenmode corrections.}
We first test the theoretical correction terms separately. This allows us to
determine whether regional mean activity and unresolved within-region
variation each contain spatial information about later tau progression. The
corresponding corrections to the quotient-scale dominant eigenmode in
Eq.~\eqref{eq:brain_perron_mode} are
\[
\widetilde r_u^{(1)}=Ru,
\qquad
\widetilde r_{uu}^{(2)},
\qquad
\widetilde r_{\theta\theta}^{(2)}
=
\sigma^2R(v\odot\chi).
\]
Initially, we take \(\chi_b=\chi\) across regions, so
\(R(v\odot\chi) \propto Rv\). 
We used the Schaefer-200 structural connectome as the quotient network and
estimated normative regional activity from FDG-PET across 47 subjects without amyloid pathology (centiloid below 20). The correction terms were
compared with the mean annualized tau-PET in 483 subjects (non-overlapping with subjects used to estimate regional activity; see Materials \& Methods,
Sec.~\ref{sec:method_neuroimaging}).

The activity-dependent corrections showed the strongest associations: the first-order mean correction
$\widetilde r_u^{(1)}$, second-order mean correction
$\widetilde r_{uu}^{(2)}$, and second-order deviation correction $Rv$
were associated with future tau accumulation with $\rho=0.437$,
$-0.438$, and $0.380$, respectively (all
$p<3.0\times10^{-8}$). By comparison, the unperturbed Perron mode
$q_0$ and mean FDG showed only weak associations
($\rho=0.182$, $p=0.010$, and $\rho=0.167$, $p=0.018$,
respectively), while FDG variance alone was not associated with future tau
accumulation ($\rho=0.073$, $p=0.306$; see Supporting Information
Fig.~S1).

\begin{figure}[!ht]
    \centering
    \includegraphics[width=\linewidth]{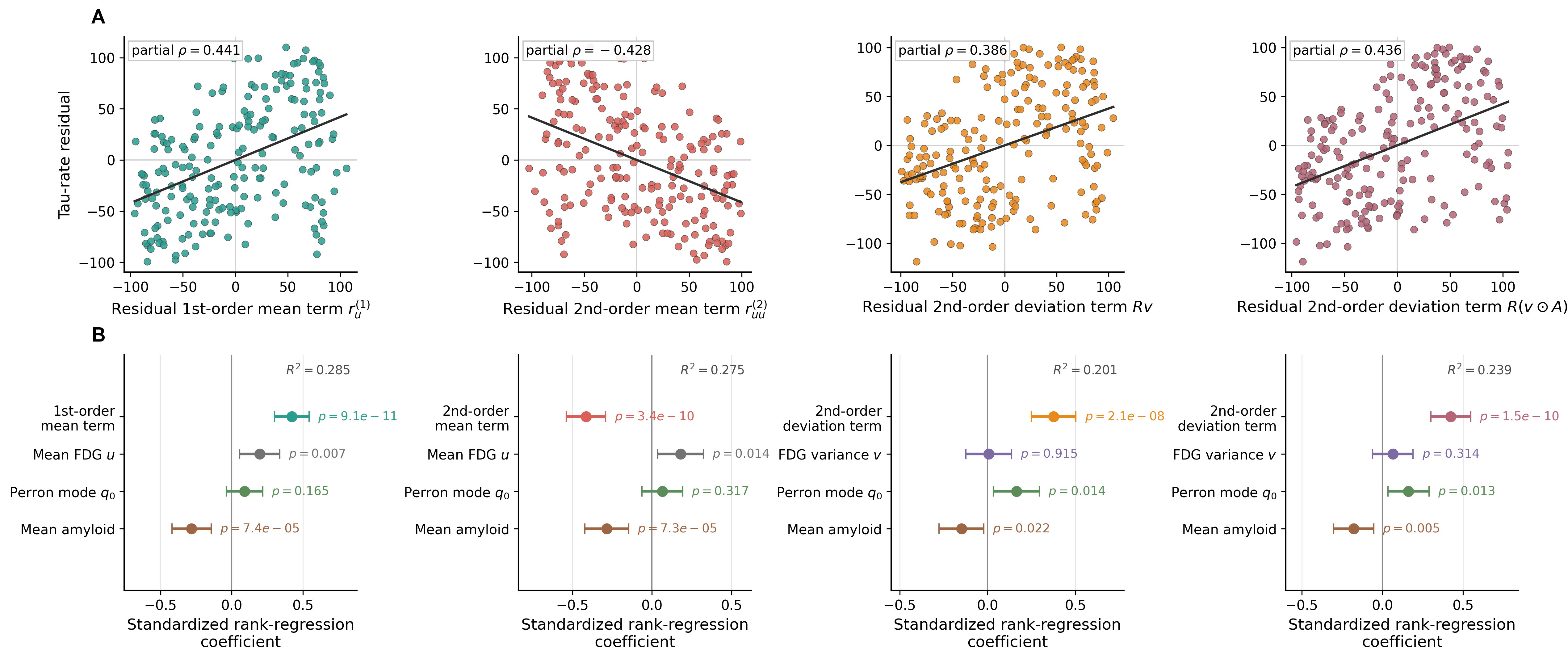}
    \caption{\textbf{Parameter-free dominant-eigenmode corrections are associated with future regional tau accumulation.}
    Mean FDG, FDG variance, and mean amyloid were estimated across 47 subjects with amyloid centiloid below 20, and future tau accumulation was averaged across 483 non-overlapping subjects. Columns show the first-order mean correction \(r_u^{(1)}\), second-order mean correction \(r_{uu}^{(2)}\), second-order deviation correction \(Rv\), and amyloid-informed second-order deviation correction \(R(v\odot A)\). (\textbf{A}) Associations between ranked predictor and tau-rate residuals after covariate adjustment. (\textbf{B}) Standardized rank-regression coefficients and 95\% confidence intervals. Mean corrections were adjusted for mean FDG, the Perron mode \(q_0\), and mean amyloid; deviation corrections were adjusted for FDG variance, \(q_0\), and mean amyloid. Each point denotes one Schaefer-200 region.}
    \label{fig:pf_parameter_free_covariate_controls}
\end{figure}

Because these corrections are constructed from the structural network and
regional imaging measures, their associations could simply reflect those
inputs. We therefore test whether each correction remains associated with
later tau progression after accounting for its constituent measures and
amyloid burden. The first- and
second-order mean corrections were adjusted for \(q_0\), mean FDG, and
mean amyloid, and the deviation correction was adjusted for \(q_0\),
FDG variance and mean amyloid (see Materials \& Methods,
Sec.~\ref{sec:method_pet_perron}). The first-order mean correction remained
positively associated with future tau accumulation (partial
\(\rho=0.441\), \(p=9.1\times10^{-11}\); standardized
\(\beta=0.421\), 95\% confidence interval \(0.300\)--\(0.542\);
\(R^2=0.285\); Fig.~\ref{fig:pf_parameter_free_covariate_controls}).
The second-order mean correction remained negatively associated
with future tau accumulation (partial \(\rho=-0.428\), \(p=3.4\times10^{-10}\);
\(\beta=-0.414\), 95\% confidence interval
\(-0.538\)--\(-0.291\); \(R^2=0.275\)), whereas the second-order
deviation correction remained positively associated (partial
\(\rho=0.386\), \(p=2.1\times10^{-8}\); \(\beta=0.376\), 95\%
confidence interval \(0.249\)--\(0.503\); \(R^2=0.201\)).

The uniform internal factor \(\chi_b=\chi\) assumes that the internal
spectral contribution has the same sign and magnitude in every region.
To allow this factor to vary regionally, we also set
\(\chi_b=s_\chi A_b\), where \(A_b\) is the mean regional amyloid centiloid and
\(s_\chi\in\{-1,+1\}\) is a common sign. This choice then gives the second-order
deviation correction \(R(v\odot A)\), up to a signed common factor. The
amyloid-informed correction remained associated with future tau after
adjustment for \(q_0\), FDG variance, and mean amyloid (partial
\(\rho=0.436\), \(p=1.5\times10^{-10}\); \(\beta=0.424\), 95\%
confidence interval \(0.300\)--\(0.547\); \(R^2=0.239\);
Fig.~\ref{fig:pf_parameter_free_covariate_controls}).

\paragraph{Full activity-dependent dominant mode.}
After examining the individual correction terms, we next test the perturbed
dominant mode itself. This determines whether the full activity-dependent
eigenmode is associated with the spatial pattern of later tau progression. We
fitted
\[
\widetilde r_0^{(1)}(\delta)
=
\widetilde q_0+\delta\widetilde r_u^{(1)}
\]
at first order, and
\[
\widetilde r_0^{(2)}(\delta,a_v;X)
=
\widetilde q_0
+\delta\widetilde r_u^{(1)}
+\delta^2\widetilde r_{uu}^{(2)}
+a_vR(v\odot X)
\]
at second order. The last term reparameterizes the theoretical
internal-deviation correction
\(\delta^2\sigma^2R(v\odot\chi)\). For the spatially uniform model,
\(X_b=1\) and \(a_v=\delta^2\sigma^2\chi\). For the amyloid-informed
model, \(X_b=A_b\) and \(a_v=\delta^2\sigma^2s_\chi\), where
\(\chi_b=s_\chi A_b\). We therefore fitted \(\delta\) at first order and
both \(\delta\) and the signed internal-deviation coefficient \(a_v\) at
second order.
To compare fitted amplitudes across corrections with different scales,
we report
$
\delta_{\mathrm{eff}}^*
=
\delta^*
 \langle|\widetilde r_u^{(1)}|\rangle_b / \langle|\widetilde q_0|\rangle_b,
$
and
$
a_{v,\mathrm{eff}}^*
=
a_v^*
\langle|R(v\odot X)|\rangle_b / \langle|\widetilde q_0|\rangle_b.
$

\begin{figure}[!ht]
    \centering
    \includegraphics[width=\linewidth]{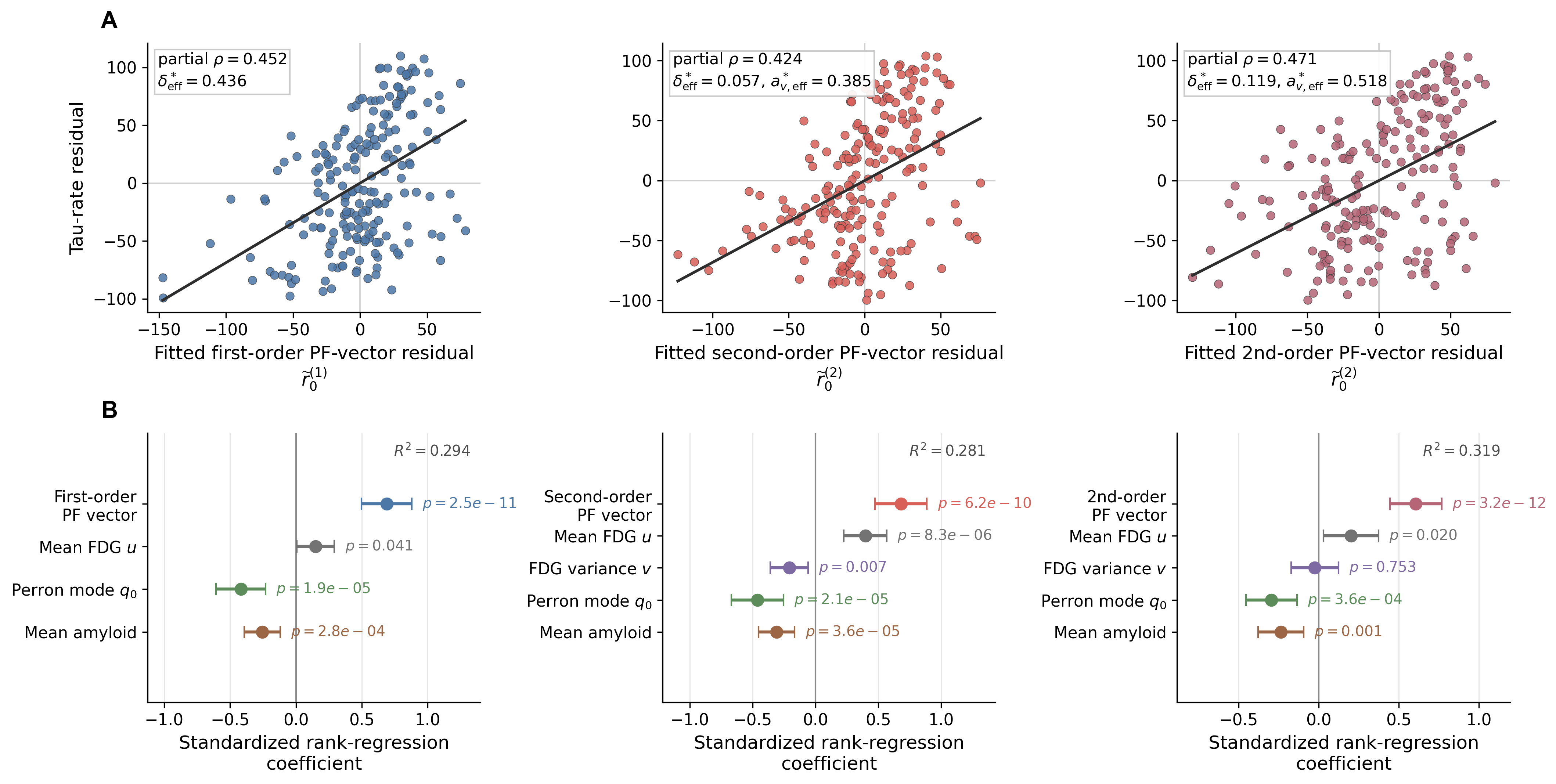}
    \caption{\textbf{Fitted activity-dependent dominant eigenmodes are associated with future regional tau accumulation.}
    Mean FDG, FDG variance, and mean amyloid were estimated across 47 subjects with centiloid below 20, while future tau accumulation was averaged across 483 non-overlapping subjects. Columns show the fitted first-order dominant eigenmode, the fitted second-order eigenmode with a uniform internal factor, and the fitted amyloid-informed second-order eigenmode. (\textbf{A}) Associations between ranked eigenmode and tau-rate residuals after covariate adjustment; annotations give partial Spearman correlations and scale-normalized fitted amplitudes. (\textbf{B}) Standardized rank-regression coefficients and 95\% confidence intervals. The first-order model was adjusted for mean FDG, the Perron mode \(q_0\), and mean amyloid; second-order models additionally included FDG variance. Each point denotes one Schaefer-200 region.}
    \label{fig:pf_fitted_covariate_controls}
\end{figure}

The optimized first-order dominant eigenmode gave \(\rho=0.434\) at
\(\delta_{\mathrm{eff}}^*=0.436\). The second-order approximation with
a uniform internal factor gave \(\rho=0.378\) at
\(\delta_{\mathrm{eff}}^*=0.057\) and
\(a_{v,\mathrm{eff}}^*=0.385\), whereas the amyloid-informed
second-order approximation gave \(\rho=0.489\) at
\(\delta_{\mathrm{eff}}^*=0.119\) and
\(a_{v,\mathrm{eff}}^*=0.518\) (Supporting Information
Fig.~S3). After adjustment for
the corresponding PET inputs, \(q_0\), and mean amyloid, the fitted
first-order, uniform second-order, and amyloid-informed second-order
dominant eigenmodes remained associated with future tau accumulation
(partial \(\rho=0.452\), \(0.424\), and \(0.471\), respectively; all
\(p<6.2\times10^{-10}\); Fig.~\ref{fig:pf_fitted_covariate_controls}).
Their standardized coefficients were \(\beta=0.688\) (95\% confidence
interval \(0.496\)--\(0.880\); \(R^2=0.294\)),
\(\beta=0.682\) (95\% confidence interval \(0.475\)--\(0.888\);
\(R^2=0.281\)), and \(\beta=0.607\) (95\% confidence interval
\(0.446\)--\(0.768\); \(R^2=0.319\)), respectively.

\paragraph{Sensitivity to the activity reference cohort.}
Because the regional activity maps were initially estimated from individuals
without substantial amyloid pathology, we repeat the analysis using the full
reference cohort to determine whether the results depend on that choice. We
used all 257 subjects regardless of their pathology levels to estimate the FDG moments and mean amyloid
(Supporting Information Sec.~C). The
partial correlations for the first-order mean, second-order mean,
second-order deviation, and amyloid-informed deviation corrections were
\(\rho=0.416\), \(-0.395\), \(0.415\), and \(0.465\), respectively
(all \(p<9.5\times10^{-9}\)). The fitted first-order, uniform
second-order, and amyloid-informed second-order dominant eigenmodes gave
partial correlations of \(\rho=0.463\), \(0.530\), and \(0.478\),
respectively (all \(p<1.5\times10^{-12}\)). Together, these regional
results support the prediction that neuronal activity changes the spatial
direction of pathological spreading. We next test the theory's complementary
prediction at the individual level by asking whether the epidemic threshold is
associated with how broadly pathology spreads.

\subsection{The first-order activity-dependent epidemic threshold is associated with tau extent}

The dominant eigenmode describes where pathological spreading first
concentrates, whereas the leading eigenvalue determines whether it can be
sustained. We next ask the complementary individual-level question: do
activity patterns that make spreading easier predict more widespread tau
pathology? We compare each subject's tau extent with the parameter-free
first-order inverse threshold. A larger inverse threshold indicates easier
spreading. For subject \(s\), we used the regional FDG map \(u_s\) from the
earliest available scan and computed
\[
\mathcal T_{s}^{-1,(1)}
=
1+\sum_b q_{0,b}^2u_{s,b},
\]
where the structural connectome was normalized to
\(\lambda_0=1\) and we set \(\delta=1\). Each FDG scan was paired with
the latest available subsequent tau scan. We quantified tau extent as
the expected number of regions assigned to the higher-tau component of
regional two-Gaussian mixture models. Because amyloid burden is strongly
associated with tau pathology, this relationship may change across stages of
amyloid accumulation. We therefore considered nested subject groups defined
by the earliest available amyloid centiloid value.

\begin{figure}[!ht]
    \centering
    \includegraphics[width=\linewidth]{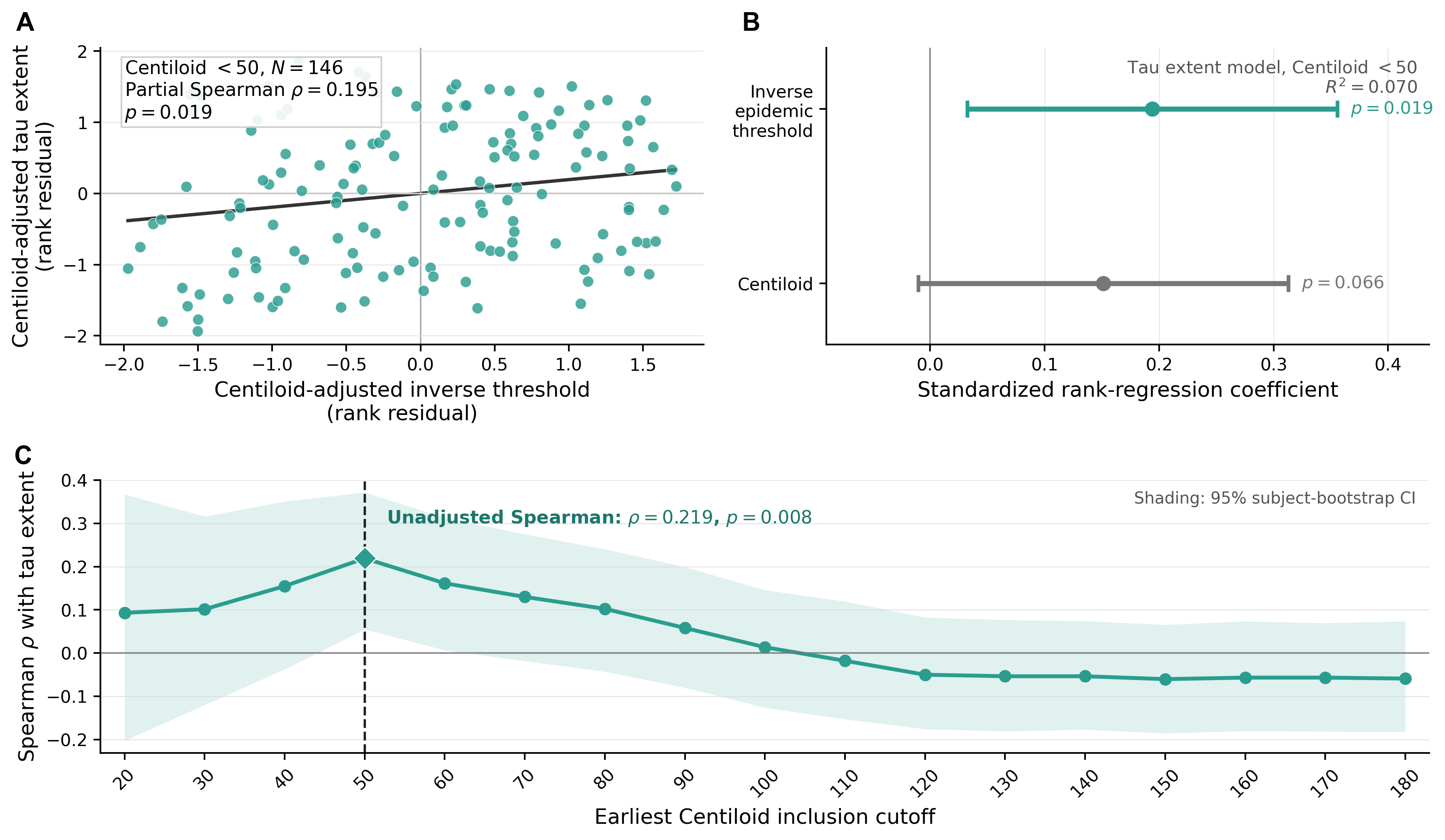}
    \caption{\textbf{The first-order inverse epidemic threshold is associated with tau extent.}
    (\textbf{A}) Centiloid-adjusted rank residuals of tau extent against the corresponding residuals of the first-order inverse epidemic threshold among subjects with earliest centiloid below 50; the line shows a least-squares fit. (\textbf{B}) Standardized rank-regression coefficients for the inverse epidemic threshold and centiloid, with 95\% confidence intervals. (\textbf{C}) Unadjusted Spearman correlation between the inverse epidemic threshold and tau extent across nested centiloid inclusion cutoffs, with 95\% subject-bootstrap confidence intervals. The dashed line marks the cutoff at the maximum observed unadjusted correlation.}
    \label{fig:first_order_threshold_tau_extent}
\end{figure}

The association between the inverse epidemic threshold and tau extent
increased as the centiloid inclusion cutoff was raised from 20 to 50,
where it reached its observed maximum
(\(N=146\), \(\rho=0.219\), \(p=0.008\), 95\% bootstrap confidence
interval \(0.055\)--\(0.371\); Fig.~\ref{fig:first_order_threshold_tau_extent}).
The association then weakened as subjects with higher centiloid values
were included and was close to zero by a cutoff of 100. The stronger
association at low-to-moderate amyloid burden is consistent with the epidemic
threshold being most relevant early in disease, when the system is approaching
the transition to sustained spreading. At later stages, pathology may already
be established, making variation in the threshold less informative. Within the
centiloid \(<50\) group, a standardized rank regression including both
the inverse threshold and centiloid retained an association for the
inverse threshold (\(\beta=0.194\), 95\% confidence interval
\(0.032\)--\(0.355\), \(p=0.019\)), whereas the amyloid centiloid coefficient
was smaller and did not reach significance (\(\beta=0.151\), 95\%
confidence interval \(-0.010\)--\(0.313\), \(p=0.066\);
\(R^2=0.070\)). Because the inclusion cutoff was selected at the
observed correlation maximum, the corresponding \(p\)-values are
nominal. We restrict this analysis to first order because individual
measurements of within-region neuronal activity variability, required
for the second-order internal contribution, were unavailable. At
low-to-moderate amyloid burden, lower predicted thresholds were therefore
associated with more extensive tau pathology, linking network stability to
individual differences in disease extent.

\section{Discussion}

Pathological proteins spread preferentially along the brain's structural
connectome, but anatomy alone does not determine whether pathology becomes
widespread or where it spreads next. Experimental studies show that neuronal
activity can promote pathological protein release and
transmission~\cite{pooler_physiological_2013,wu_neuronal_2016,pooler_amyloid_2015}.
Our theoretical analysis shows that neuronal activity shifts both the epidemic
threshold and the initial direction of spread along structural pathways. In
longitudinal human imaging data, these activity-dependent spectral changes are
associated with subsequent tau progression.

\subsection{Activity reshapes the onset and spatial course of pathological spreading}

Existing connectome models explain how anatomical connectivity constrains the
propagation of established pathology, but they generally determine transmission
from structural connections
alone~\cite{raj_network_2012,fornari_prion-like_2019,yang_longitudinal_2019,yang_longitudinal_2021,vogel_spread_2020,schafer_network_2020,schafer_predicting_2021,chaggar_personalised_2025}.
Our earlier work coupled protein spreading to neuronal oscillations
and showed that heterogeneous activity can lower the threshold for toxic
spreading and break symmetry in the spreading
pattern~\cite{alexandersen_neuronal_2024}. A complementary transport model
weighted tau movement by regional neuronal activity or amyloid burden and used
this mechanism to explain the selective vulnerability of the entorhinal cortex
to early tau seeding~\cite{alexandersen_neuronal_2026}. Here, we take a different
approach. Rather than coupling specific neuronal and protein dynamics or
weighting tau transport using a particular biological map, we derive general
spectral principles for how activity reshapes spreading on a fixed network. The
framework approximates the epidemic threshold and Perron mode, separates
regional from within-region activity effects, and connects both quantities to
longitudinal tau progression.

The threshold is most sensitive to activity in regions with large entries in
the structural Perron mode, which dominate the connectome's spreading pattern.
Activity-modulating interventions may therefore need to account for network
position.

Brain imaging averages activity across many neurons within each region. Our
multiscale analysis shows that variation hidden by this averaging can still
influence brain-wide spreading through within-region network structure. The
decomposition approximates this influence without resolving every neuron.
Discrepancies between regional imaging measurements and disease progression may
therefore reflect missing spatial scales rather than an absence of
activity-dependent effects.

\subsection{Activity--network interactions predict human tau progression}

The longitudinal human imaging analyses show that activity-dependent changes
to spreading are not only a theoretical possibility. Coupling regional
activity to structural connectivity captured later tau accumulation better
than either quantity alone, showing that their interaction helps determine the
spatial course of disease. An amyloid-informed estimate of unresolved activity
variation also remained associated with later tau progression after accounting
for regional amyloid, complementing work linking amyloid-related
changes in neuronal activity and functional connectivity to tau
progression~\cite{giorgio_amyloid_2024,hojjati_inter-network_2025,roemer-cassiano_amyloid-associated_2025}.
Amyloid may therefore provide additional information about activity-dependent
spreading.

At low-to-moderate amyloid burden, lower predicted thresholds were associated
with more extensive tau pathology, consistent with the threshold being most
relevant as pathology becomes widespread. Together with the regional analysis,
this links neuronal activity to both the onset and direction of spread and
extends our earlier findings that heterogeneous activity can destabilize
healthy dynamics and break spatial symmetry~\cite{alexandersen_neuronal_2024},
and that glucose metabolism and amyloid patterns can bias where tau
accumulates~\cite{alexandersen_neuronal_2026}.

\subsection{Methodological Considerations}

Our framework is designed to describe the early invasion of pathology, when a
small seed either dies out or begins to spread. It therefore does not capture
later protein aggregation, neuronal degeneration, or structural damage. The
susceptible--infected--susceptible dynamics likewise represent transmission at
a coarse level rather than literal protein clearance or recovery. These
simplifications make the threshold and Perron mode analytically tractable, but
restrict the framework to the earliest stage of spreading.

The multiscale analysis makes related simplifying assumptions. Idealized block
structure and localized internal modes allow us to isolate how within-region
variation affects brain-wide spreading, although real brain networks will
satisfy these assumptions only approximately.

The imaging measures also provide indirect proxies for model quantities.
Glucose metabolism does not directly measure neuronal firing, and
across-subject regional variance only approximates activity variation within
brain regions. The human data therefore test the regional predictions more
directly than the proposed fine-scale mechanism.

\subsection{Future Directions}

Direct measurements of within-region activity variation are needed to test the
multiscale mechanism. Voxel-wise glucose metabolism, local functional-imaging
variability, or source-resolved electrophysiology could measure the fine-scale
patterns that influence spreading. Individualized connectomes could test how
person-specific anatomy modifies these effects. Functional connectivity may
capture state-dependent transmission, but should modulate structural pathways
rather than replace them.

The framework also defines a control problem. Reducing activity in regions with
large \(q_{0,i}^2\) most efficiently raises the threshold, while \(R\) identifies
how targeted interventions redirect early spread. These quantities may guide
where activity-modulating interventions, including neuronal stimulation, could
most effectively slow or prevent progression. They also connect the framework to epidemic
resource allocation~\cite{preciado_optimal_2014,nowzari_analysis_2016} and
network control theory~\cite{gu_controllability_2015}.

\subsection{Conclusion}

The same spectral principles governing epidemics on networks also shape how
Alzheimer's disease begins and spreads through the brain. Structural
connectivity defines the routes of spread, while neuronal activity changes the
threshold and spatial direction. Activity variation hidden by regional
averaging can also influence brain-wide progression. Their correspondence with
longitudinal tau pathology implicates neuronal activity as a driver of disease
progression and motivates interventions to slow or prevent pathological spread.
More broadly, this work shows how the physics of epidemic spreading can offer
new insight into neurodegenerative disease.

\section{Methods}
\subsection{Network and state variables}

We consider a weighted, directed graph without self-loops captured by the weighted adjacency matrix $W$, with elements $W_{ij}$. The entries of this matrix, $W_{ij}$, capture interactions from node $j$ to node $i$.

We study two discrete time Markov chain (DTMC) processes occurring on this graph: a spreading process and an activity process. The spreading process describes the transmission of pathology between nodes, while the activity process describes neuronal firing dynamics. Each node is therefore characterized by both an infection state and an activity state.

\subsection{SIS and QAQ dynamics}

\paragraph{SIS model.}
We consider a two-state spreading model where a node can be in one of two possible states: susceptible (S) or infected (I). We assume that the probability of going from infected to susceptible is independent of the states of its neighbours,
\begin{equation}
    P( I_i \rightarrow S_i) = \zeta,
\end{equation}
where $\zeta \geq 0$ is the transition rate from infected to susceptible.
Then, the rate of infection increases linearly with infected neighbours giving
\begin{equation}
    P(S_i \rightarrow I_i) =  \beta \sum_{j=1}^N W_{ij} X^{(I)}_j,
\end{equation}
where $\beta \geq 0$ is the rate of infection.

\paragraph{QAQ model.}
Consider a two-state neuron model where a neuron can be in one of two possible states: Active (A) or Quiet (Q). We assume that the probability of entering the refractory period is independent of input:
\begin{equation}
    P( A_i \rightarrow Q_i) = \gamma.
\end{equation}
We assume stochasticity in the threshold for firing $I > I_c + \varepsilon$ where $\varepsilon \sim \mathcal{N} (0,\sigma^2_{\varepsilon})$. The neuron fires if $I_i > I_c + \varepsilon$ and so
\begin{align}
    P(Q_i \rightarrow A_i) &= P (I_i > I_c + \varepsilon) \\
     &= P (\varepsilon < I_i - I_c) \\
     &= \Phi\left(\frac{I_i - I_c}{\sigma_\varepsilon}\right),
\end{align}
where $\Phi$ is the CDF of the standard normal distribution. The input to a node $i$ is given by
\[
I_i = \alpha \sum_{j=1}^N W_{ij} X^{(A)}_j,
\]
where $\alpha > 0$ is the global coupling strength.

\subsection{Coupled SISQAQ model}\label{sec:method_MMCA}

We now account for the possibility that the state of infection of a node affects its threshold for firing, and that the rate of infection is different when the neuron is firing~\cite{wu_neuronal_2016}. 

\paragraph{Activity subsystem.}
The probability of entering the refractory period is independent of infection:
\begin{equation}
    P( A_i \rightarrow Q_i) = \gamma.
\end{equation}
In contrast, the threshold for firing is affected by infection, such that the mean is shifted $I_{c,i} = I^0_c - \kappa X^{(I)}_i$ and the variance in the threshold is independent of the infection state $\varepsilon \sim \mathcal{N} (0,\sigma^2_{\varepsilon})$. The parameter $\kappa$ could be positive or negative depending on how the infection affects the firing. 
\begin{align}
    P(Q_i \rightarrow A_i) &= P (I_i > I_{c,i} + \varepsilon) \\
     &= P (\varepsilon < I_i - I_{c,i}) \\
     &= \Phi(I_i - I_{c,i}; \mu = 0, \sigma^2 = \sigma^2_{\varepsilon}),
\end{align}
where $\Phi$ is the CDF of the normal distribution. The input to a node $i$ is given as before
\[
I_i = \alpha \sum_{j=1}^N W_{ij} X^{(A)}_j.
\]

\paragraph{Spreading subsystem.}
We introduce a timescale separation parameter $\varepsilon > 0$ and consider regimes where the spreading process occurs on a slower timescale $\varepsilon \ll 1$ than the activity process. We assume that the probability of going from infected to susceptible is independent of the states of its neighbours and activity:
\begin{equation}
    P( I_i \rightarrow S_i) = \varepsilon \zeta
\end{equation}
and that the rate of infection is modified by the source node's activity state
\begin{equation}
    P(S_i \rightarrow I_i) = \varepsilon \beta \sum_{j=1}^N W_{ij} (1 + \delta X^{(A)}_j) X^{(I)}_j,
\end{equation}
where $\beta \geq 0$ is the rate of infection and $\delta \geq -1$ determines how much neuronal activity impacts infection. When $\delta>0$, neuronal activity increases the chance of transmission, and \emph{vice versa} for $\delta<0$.

\subsection{Numerical verification of the threshold and dominant eigenmode}\label{sec:method_numerical}

We tested the spectral predictions on weighted, undirected synthetic
networks and in stochastic simulations of the coupled SISQAQ process. All
network matrices were checked for symmetry and nonnegative entries and
rescaled so that \(\rho(W)=1\). For a given network, the disease-free activity
probabilities were obtained by iterating the QAQ mean-field map
\[
a_i^{(t+1)}
=
a_i^{(t)}
+
\left(1-a_i^{(t)}\right)
\Phi\!\left(
\frac{\alpha(Wa^{(t)})_i-I_c^0}
{\sqrt{\sigma_\varepsilon^2+
\alpha^2[W^{\circ 2}(a^{(t)}\odot(1-a^{(t)}))]_i}}
\right)
-
\gamma_i a_i^{(t)}
\]
from \(a_i^{(0)}=0.5\) until the maximum absolute change was below
\(10^{-12}\). Here \(W^{\circ 2}\) is the element-wise square of \(W\).
This calculation accounts for fluctuations in the Bernoulli activity
states entering each node and was used only to obtain the disease-free
activity vector for the spectral calculations.

\paragraph{Synthetic network families and spectral calculations.}
The cross-network comparison used five networks with \(N=200\) nodes. The
all-to-all network had \(W_{ij}=1/(N-1)\) for \(i\ne j\). The rank-one
network was \(W=cxx^\top\), where the positive entries of \(x\) were
sampled from a shifted Pareto distribution with shape parameter \(3\).
The weighted Erd\H{o}s--R\'enyi network had connection probability
\(p=0.05\)~\cite{erdos59a}, and the scale-free network was generated by
Barab\'asi--Albert preferential attachment with two edges added per new
node~\cite{barabasi_emergence_1999}. Existing edges in both random
networks received weights sampled uniformly from \(0.25\) to \(1\).
Finally, the two-block network contained two complete blocks of 100 nodes
with equal within-block spectral radii and dense inter-block edges whose
unscaled weights were sampled uniformly from \(0.25\) to \(1\) and then
multiplied by \(0.002\). The common QAQ parameters were
\(\gamma=0.34\), \(I_c^0=0.73\), \(\sigma_\varepsilon=0.33\),
\(\kappa=0\), and \(\alpha=1.05\). The two-block network instead used
\(\gamma_1=0.175\) and \(\gamma_2=0.90\) to produce different activity
levels in the two blocks.

For each activity pattern, we formed
\(A(\delta)=\operatorname{diag}(\mathbf 1+\delta a)\). Because \(W\) was
symmetric, the exact leading eigenvalue of \(WA(\delta)\) was calculated
from the symmetric similar matrix
\(A(\delta)^{1/2}WA(\delta)^{1/2}\). Its right Perron mode was recovered
by multiplying the corresponding eigenvector by \(A(\delta)^{-1/2}\),
orienting it positively, and imposing \(q_0^\top r_0(\delta)=1\). The
first- and second-order eigenvalue and mode approximations were
calculated from Eqs.~\eqref{eq:symmetric_spectral_radius}--\eqref{eq:perron_vector_second_order}.
Thresholds were expressed as
\(\beta_c=\zeta/\rho(WA)\), with \(\zeta=0.5\), and divided by their value
at \(\delta=0\) for the relative-threshold plots. We evaluated 121
equally spaced values of \(\delta\) from \(-0.75\) to \(3\), except for
the two-block network, for which the range was \(-0.5\) to \(1\).
Perron-mode errors were calculated as the relative Euclidean error
\(\|r_{\mathrm{exact}}-r_{\mathrm{approx}}\|_2/
\|r_{\mathrm{exact}}\|_2\).

\paragraph{Quasi-stationary threshold estimates.}
Stochastic thresholds were estimated by scanning \(\beta\) in
quasi-stationary SISQAQ simulations~\cite{de_oliveira_how_2005,
ferreira_epidemic_2012}. The QAQ process was first equilibrated with no
infection, after which infection was seeded uniformly in \(1\%\) of the
nodes. We implemented the timescale separation by performing
\(m_{\mathrm{QAQ}}=10\) QAQ updates before each SIS update. For every
\(\beta\), 16 independent runs were equilibrated for
\(T_{\mathrm{eq}}=60\) SIS steps and observed for
\(T_{\mathrm{obs}}=120\) further steps. The quasi-stationary process
stored 64 non-extinct joint activity--infection states. An extinct run
was restarted from a randomly chosen stored state; after the archive was
full, its contents were refreshed by replacing a randomly chosen state
with probability \(0.05\) at each non-extinct step.

For each \(\beta\), moments were pooled across runs and used to calculate
\[
\chi_{\mathrm{QS}}
=
\frac{N\left(\langle\rho_I^2\rangle-\langle\rho_I\rangle^2\right)}
{\langle\rho_I\rangle},
\]
where \(\rho_I\) is infected prevalence. We first evaluated 25
transmission rates between \(0.001\) and \(1.5\), selected the first
substantial local maximum of \(\chi_{\mathrm{QS}}\), and refined its
neighborhood on a denser grid. The empirical threshold was the vertex of
a concave quadratic fitted locally to the peak; when the fitted vertex
fell outside the local interval, the largest observed value was used.
Numerically unstable scans were repeated with at least 24 runs and 150
observation steps.

\paragraph{Dominant eigenmode validation.}
The dominant eigenmode was tested on independent realizations of the
same five network families. For this analysis, the Erd\H{o}s--R\'enyi
connection probability was \(0.025\), the scale-free network was a
Chung--Lu graph with Pareto exponent \(3.5\)~\cite{chung_connected_2002},
and the two-block inter-block weight scale was \(5\times10^{-4}\).
The QAQ parameters were \(\gamma=0.4\), \(I_c^0=0.83\),
\(\sigma_\varepsilon=0.4\), and \(\alpha=1.25\), with
\(\gamma_1=0.30\) and \(\gamma_2=0.98\) in the two-block network.
Exact and perturbative modes were compared over 121 logarithmically
spaced values of \(\delta\) from \(10^{-3}\) to \(3\).

To test the mode against the stochastic dynamics, we generated 24 QAQ
activity paths after 60 equilibration steps and again used 10 QAQ updates
per SIS update. At \(\delta=1\), we simulated 384 independent infection
runs for 32 SIS steps at \(\beta=1.08\beta_c\), with infection seeded
uniformly in \(1\%\) of nodes. The empirical early-growth profile was the
mean node occupancy over a four-step window ending at the latest time
\(t\ge7\) for which mean prevalence was between \(5\times10^{-4}\) and
\(0.06\). If no time met these criteria, we used the time closest to
prevalence \(0.03\). Exact, approximate, and stochastic profiles were
normalized to sum to one and compared using cosine similarity and
Spearman correlation.

\paragraph{Heterogeneous QAQ parameter sweeps.}
We varied one parameter at a time on a fixed weighted
Erd\H{o}s--R\'enyi realization with \(N=200\), connection probability
\(p=0.05\), and \(\rho(W)=1\). Existing edges received symmetric random
weights generated from values between \(0.25\) and \(1\) before the
spectral normalization. Baseline parameters were
\(\gamma=0.3\), \(I_c^0=0.5\), \(\sigma_\varepsilon=0.4\),
\(\kappa=0\), \(\zeta=0.5\), \(\varepsilon=1\), and \(\alpha=1\);
\(\delta=1\) was fixed outside the \(\delta\) sweep. Nine equally spaced
values were used for each parameter:
\(\delta\in[-0.675,1.5]\), \(\alpha\in[0.5,2]\),
\(I_c^0\in[0.2,0.8]\), \(\gamma\in[0.1,0.6]\), and
\(\sigma_\varepsilon\in[0.15,0.5]\).

Exact, first-, and second-order thresholds and exact, zeroth-, first-,
and second-order Perron modes were additionally evaluated at 101
points over each parameter range to obtain the smooth curves and
approximation errors. Perron modes were normalized to sum to one
before their relative Euclidean errors were calculated. The
quasi-stationary thresholds were estimated at the nine simulated values
per sweep using the settings described above. For the relative-threshold
comparisons, every theoretical and simulated series was divided by its
own value at \(\delta=0\), \(\alpha=1\), \(I_c^0=0.5\),
\(\gamma=0.3\), or \(\sigma_\varepsilon=0.4\), as appropriate, using
linear interpolation between simulated points when the reference value
was not sampled directly.

At each of the 45 parameter settings, we generated 96 QAQ activity paths
after 60 equilibration steps, used 10 QAQ updates per SIS update, and
simulated 1,536 early-growth infection runs for 32 SIS steps at
\(\beta=1.08\beta_c\), with infection seeded uniformly in \(1\%\) of
nodes. With 200 nodes at each setting, the pooled comparison comprised
9,000 node--setting pairs. Profiles were selected and normalized as in the cross-network
dominant eigenmode validation. For the pooled Perron-mode parity plot,
each predicted and simulated profile was divided by its mean across
nodes, so one denotes the network-average regional growth weight. At
each parameter setting, we regressed the node-wise SISQAQ profile on each
predicted Perron mode with an intercept and recorded \(R^2\), the
fraction of regional early-growth variation explained by the predicted
spatial pattern.

\subsection{Numerical verification of the multiscale decomposition}
\label{sec:method_multiscale_validation}

We constructed a synthetic equitable network designed to activate both
the quotient-scale and internal-mode corrections. The network contained
four blocks of 50 nodes, each forming a weighted ring with a known
localized cosine eigenmode. The respective nearest-neighbor weights were
\((0.30,0.24,0.16,0.32)\). The unequal ring weights gave distinct
internal eigenvalues, while the fourth ring produced a slowly decaying
centered mode with eigenvalue \(0.971\) after normalizing
\(\rho(W)=1\), corresponding to a spectral gap of \(0.029\).  Every pair of nodes belonging to different blocks was
connected with equal weight. If \(b\ne c\), each edge between
blocks \(b\) and \(c\) had weight \(B_{bc}/50\), where the block-to-block
strength matrix was
\[
B
=
0.40
\begin{pmatrix}
0 & 0.08 & 0.03 & 0.05\\
0.08 & 0 & 0.06 & 0.02\\
0.03 & 0.06 & 0 & 0.07\\
0.05 & 0.02 & 0.07 & 0
\end{pmatrix}.
\]
The full network and its quotient matrix were rescaled together so that
\(\rho(W)=1\). The centered internal mode in each ring was
\(s_{b,i}=\sqrt{2/50}\cos(2\pi i/50)\). We prescribed block-average
activities \((0.2,0.4,0.6,0.8)\) and added a cosine deviation of peak
amplitude \(0.1\) within every block. Node-specific \(\gamma_i\) values
were then calculated so that this designed activity pattern was an exact
fixed point of the disease-free QAQ map:
\[
\gamma_i
=
\frac{1-a_i}{a_i}
\Phi\!\left(
\frac{\alpha(Wa)_i-I_c^0}
{\sqrt{\sigma_\varepsilon^2+
\alpha^2[W^{\circ 2}(a\odot(1-a))]_i}}
\right).
\]
The remaining parameters were
\(I_c^0=1\), \(\sigma_\varepsilon=0.6\), \(\kappa=0\),
\(\zeta=0.5\), \(\varepsilon=1\), and \(\alpha=1\).

The exact full-matrix threshold and Perron mode, the blockwise
first-order approximation, and the complete second-order approximation
were evaluated at 121 values of \(\delta\) from \(-0.2\) to \(0.2\).
The blockwise terms were also compared directly with the corresponding
full-matrix perturbation terms as an implementation check. Threshold
error was measured relative to the exact threshold, and Perron-mode
error was the Euclidean distance between unit-normalized exact and
approximate vectors.

For the stochastic validation, 24 QAQ paths were equilibrated for 60
steps. Quasi-stationary thresholds were estimated at
\(\delta=-0.20,-0.10,0,0.05,0.10,0.15,\) and \(0.20\), using 20 runs,
60 equilibration steps, 120 observation steps, 13 transmission rates,
and the same archive and QAQ-to-SIS update settings described above. At
\(\delta=0.1\), the early-growth infection profile was averaged over
4096 runs of 32 SIS steps at \(1.08\) times the empirical threshold.
Infection was seeded in \(1\%\) of nodes with probabilities proportional
to the unperturbed Perron mode, and the early-growth window was selected
as in the cross-network validation.

\subsection{Neuroimaging data and structural connectome}
\label{sec:method_neuroimaging}

We analyzed deidentified FDG-, amyloid-, and tau-PET data from the
Alzheimer's Disease Neuroimaging Initiative (ADNI), the Harvard Aging
Brain Study (HABS), and the Anti-Amyloid Treatment in Asymptomatic
Alzheimer's Disease study (A4)~\cite{petersen_alzheimers_2010,
dagley_harvard_2017,sperling_trial_2023}. The present analyses used preprocessed
regional standardized uptake value ratios (SUVRs) represented in the
Schaefer-200 cortical atlas~\cite{schaefer_local-global_2018}. For each
modality, regional columns were required to follow the same parcel order
from 1 to 200.

The quotient network was a group structural connectome derived from the
Human Connectome Project and represented using the same 200 Schaefer
parcels~\cite{van_essen_wu-minn_2013,schaefer_local-global_2018}. Each node
corresponded to one parcel, so \(B=200\) in every neuroimaging analysis. The
\(200\times200\) matrix was symmetric and nonnegative, had a zero
diagonal, and was rescaled to leading eigenvalue \(\lambda_0=1\). Its
eigenvectors were oriented consistently, with the Perron mode \(q_0\) chosen
positive and normalized to unit Euclidean norm. The same connectome and parcel
ordering were used in every PET analysis.

\paragraph{Normative FDG maps.}
To construct the population activity maps used in the regional
eigenmode analyses, we formed a joint HABS+ADNI cohort with complete
200-region FDG and amyloid maps at the earliest available scans and a
complete tau scan strictly after both baseline PET scans. We excluded
subjects with nonfinite regional data or mean tau SUVR greater than or
equal to 6, leaving 257 subjects. The centiloid value was recovered from
the selected amyloid scan. The primary reference group comprised the 47
subjects with centiloid below 20; analyses using all 257 subjects were
performed as a sensitivity analysis. For each reference group, the
regional mean activity and activity-variability maps were
\[
u_b=\frac{1}{n}\sum_{s=1}^{n}F_{s,b},
\qquad
v_b=\frac{1}{n-1}\sum_{s=1}^{n}(F_{s,b}-u_b)^2,
\]
where \(F_{s,b}\) is raw regional FDG SUVR. No subject-wise or regional
scaling was applied in these analyses. When amyloid was used to specify
heterogeneous internal spectral factors, we also calculated
\(A_b=n^{-1}\sum_s A_{s,b}\) from the raw regional amyloid values in
the same reference subjects.

\paragraph{Future regional tau accumulation.}
The regional outcome was constructed from the combined A4+ADNI
dataset. We retained visits with complete 200-region tau and
amyloid maps, sorted visits by tau date, and used the first consecutive
tau interval for each subject. To keep the FDG reference and tau outcome
independent, all 101 subject identifiers shared with the HABS+ADNI
reference cohort were excluded. This left 483 subjects. For subject
\(s\), the annualized regional change was
\[
\dot T_{s,b}
=
\frac{T_{s,b}^{(2)}-T_{s,b}^{(1)}}{\Delta t_s},
\]
where \(\Delta t_s\) is measured in years. The outcome used in the
regional analyses was the mean of \(\dot T_{s,b}\) across these 483
subjects for each of the 200 regions.

\subsection{PET analysis of the dominant eigenmode}
\label{sec:method_pet_perron}

We applied the operator \(R\) in
Eq.~\eqref{eq:quotient_response_operator} to the structural connectome
and the population PET quantities. The parameter-free regional
quantities were the unperturbed Perron mode \(q_0\), the mean and
variance FDG inputs \(u\) and \(v\), the first-order mean correction
\(\widetilde r_u^{(1)}=Ru\), the complete second-order mean correction
\(\widetilde r_{uu}^{(2)}\) from
Eq.~\eqref{eq:second_order_mean_perron_response}, and the second-order
deviation correction \(Rv\). Common positive scalar factors were
omitted when they did not change regional rank ordering. For the
unadjusted analyses, each input or correction was compared with future
regional tau accumulation using a two-sided Spearman correlation across
the 200 regions.

We also reconstructed complete perturbative dominant eigenmodes. At
first order we used
\[
\widetilde r_0^{(1)}(\delta)=q_0+\delta Ru.
\]
At second order we used
\[
\widetilde r_0^{(2)}(\delta,a_v)
=
q_0+\delta Ru+\delta^2\widetilde r_{uu}^{(2)}
+
a_vRv,
\]
where \(a_v=\delta^2\sigma^2\chi\) is the signed internal-deviation
coefficient under the spatially uniform assumption \(\chi_b=\chi\).
The first-order \(\delta\) was selected from 5001 values over the
admissible part of \([-10,10]\). For the second-order mode, we first
evaluated a \(501\times501\) grid over \(\delta\) and \(a_v\), followed
by a \(201\times201\) local refinement. For each \(\delta\), the
admissible range of \(a_v\) was obtained directly from the linear
inequalities requiring every entry of the complete mode to remain
positive. In both cases the objective was to maximize Spearman
correlation with the 200-region tau-rate map. We also required
\(1+\delta u_b>0\) for every region.

To specify heterogeneous internal spectral factors, we set
\(\chi_b=s_\chi A_b\), where \(A_b\) is mean regional amyloid and
\(s_\chi\in\{-1,+1\}\) is the global sign relating amyloid to the
internal spectral factor. The second-order deviation correction then
becomes \(R(v\odot A)\), up to a common signed factor. The fitted
coefficient is \(a_v=\delta^2\sigma^2s_\chi\), so reversing every
regional factor is absorbed by its sign. The complete second-order
dominant eigenmode was refitted with this correction using the same
search and positivity constraints.

Because the fitted parameters multiply corrections with different
scales, we additionally report scale-normalized effective amplitudes,
\[
\delta_{\mathrm{eff}}^*
=
\delta^*
\frac{\langle|\widetilde r_u^{(1)}|\rangle_b}
{\langle|q_0|\rangle_b},
\qquad
a_{v,\mathrm{eff}}^*
=
a_v^*
\frac{\langle|R(v\odot X)|\rangle_b}
{\langle|q_0|\rangle_b},
\]
where \(X=1\) for a uniform internal factor and \(X=A\) for the
amyloid-informed factor. These transformations were applied only for
reporting and did not alter parameter fitting or the reconstructed
dominant eigenmodes.

To determine whether the dominant-eigenmode corrections contained
information beyond their regional inputs, all variables were
rank-transformed before adjustment. Models for
\(\widetilde r_u^{(1)}\) and \(\widetilde r_{uu}^{(2)}\) included mean
FDG, \(q_0\), and mean amyloid as covariates. Models for \(Rv\) and
\(R(v\odot A)\) included FDG variance, \(q_0\), and mean amyloid. The
fitted first-order dominant eigenmode was adjusted for mean FDG,
\(q_0\), and mean amyloid, while both fitted second-order eigenmodes
were adjusted for mean FDG, FDG variance, \(q_0\), and mean amyloid.
Partial Spearman correlations were calculated by correlating the
predictor and tau-rate residuals after linear adjustment for the same
ranked covariates. We also fitted ordinary least-squares models to
standardized ranks and report standardized coefficients, 95\%
confidence intervals, two-sided \(p\)-values, and model \(R^2\). The
fitted parameters and their correlations are explanatory, and the
reported correlation \(p\)-values do not account for parameter
selection on the same outcome. All regional tests were uncorrected.
Because regions are spatially dependent, and because the optimized
dominant eigenmodes were fitted to the same regional outcome, the
associated \(p\)-values are nominal.

\subsection{Subject-level first-order threshold and tau extent}
\label{sec:method_subject_threshold}

For each HABS+ADNI subject, we selected the earliest complete FDG scan,
the earliest complete amyloid scan with a finite centiloid value, and the
latest complete tau scan obtained on or after the FDG scan. This procedure yielded
263 subjects. FDG was min--max scaled once across all retained subjects
and all 200 regions; individual scans were not scaled separately. The
first-order inverse epidemic threshold was
\[
\mathcal T_s^{-1,(1)}
=
1+\sum_{b=1}^{200}q_{0,b}^2u_{s,b},
\]
where \(u_{s,b}\) is the globally scaled FDG value and \(\delta=1\).
Because \(\sum_bq_{0,b}^2=1\), any common positive affine rescaling of
FDG preserves subject ordering under this score.

Tau extent was defined using separate two-component Gaussian mixture
models for each Schaefer region. The mixtures were trained on the
earliest complete tau scan from each unique subject pooled across A4,
ADNI, and HABS. Cross-file duplicates were removed by subject identifier,
leaving 940 subjects. Each model used full one-dimensional covariance,
20 random initializations, a maximum of 1000 iterations, and covariance
regularization \(10^{-6}\). The component with the larger mean was
defined as the high-tau component. For subject \(s\), tau extent was the
sum across regions of the posterior probabilities of belonging to this
component, and can therefore be interpreted as the expected number of
high-tau regions.

We calculated the Spearman association between the inverse threshold and
tau extent in nested samples defined by earliest centiloid cutoffs from
20 to 180 in increments of 10. Percentile 95\% confidence intervals were
calculated from 2000 subject-level bootstrap samples. The cutoff with the
largest observed unadjusted correlation was used for the displayed
scatter and the adjusted analysis. Within that sample, tau extent,
inverse threshold, and centiloid were rank-transformed and standardized,
and an ordinary least-squares model was fitted with tau extent as the
outcome and inverse threshold and centiloid as predictors. We report
standardized coefficients, parametric 95\% confidence intervals, and
two-sided \(p\)-values. Because the cutoff was selected from the same
data, these inferential quantities are nominal.
\printbibliography

\end{document}